\documentclass{ws-procs11x85}

\makeindex
\begin{document}

\title{HEAVY QUARKONIUM\footnotemark\\[-0.3cm]}
%, PRODUCTION AND SPECTROSCOPY}

\author{T. SKWARNICKI}

\address{Department of Physics, 
         Syracuse University, NY 13244, USA\\E-mail: tomasz@physics.syr.edu\\[-0.2cm]}

%\author{A. N. OTHER}
%
%\address{Department of Theoretical Physics, 1 Keble
%Road, Oxford OX1 3NP, England\\E-mail: other@tp.ox.uk}

%%%%%%%%%%%%%%%%%%%%%%%%%%%%%%%%%%%%%%%%%%%%%%%%%%%%%%%%%%%%%%%%%%%%%%%%%
% You may repeat \author \address as often as necessary             %
%%%%%%%%%%%%%%%%%%%%%%%%%%%%%%%%%%%%%%%%%%%%%%%%%%%%%%%%%%%%%%%%%%%%%%%%%

\twocolumn[\maketitle\abstract{ 
I review heavy quarkonium physics in view of recent 
experimental results. 
In particular, I discuss new results on spin singlet states,
photon and hadronic transitions, $D-$states and discovery
of the yet unexplained narrow $X(3872)$ state.
}\vspace{0.5cm}]

\def\ee{e^+e^-}
\def\mm{\mu^+\mu^-}
\def\cc{c\bar c}
\def\bb{b\bar b}
\def\DD{D\bar D}
\def\BB{B\bar B}
\def\Y{\Upsilon}
\def\etal{{\it et al.}}
\def\myref#1{\hbox{\raisebox{-3.5pt}{{\large\protect\cite{#1}}}}}
\def\Myref#1{\hbox{\raisebox{-4pt}{{\Large\protect\cite{#1}}}}}
\def\mcite#1{\protect\cite{#1}}
\def\B{{\cal B}}
\def\JD{J_D}
\def\JTwo{J_2}
\def\JOne{J_1}
% need level diagram, mention transition selection rules.

\baselineskip=13.07pt

\begin{table*}[bthp]
\caption{
Some properties of various ``onia''.\label{tab:onia}}
%\vskip-0.5cm
%\vspace{0.2cm}
\begin{center}
\def\1#1#2#3{\multicolumn{#1}{#2}{#3}}
\tablefont{
\begin{tabular}{|c|c|c|c|c|r|c|l|c|c|}
\hline
\1{3}{|c|}{FORCES} &			
System	&         
\1{3}{c|}{Ground triplet state $1^3S_1$} &			
$\left(\frac{v}{c}\right)^2$ &
\1{2}{c|}{
\begin{minipage}{2.5cm}
Number of states \linebreak 
below dissociation \linebreak
energy
\end{minipage}
} \\
%
%
%Number of states} \\ %below dissociation energy 
%\1{3}{|c|}{\quad} &			
%	&         
%\1{3}{c|}{\quad} &			
%        &
%\1{2}{c|}{below dissociation energy} \\ 
\cline{1-3} \cline{5-7} \cline{9-10}
binding	& \1{2}{c|}{decay} &		
        &
Name    & $\Gamma$ (MeV) & Mass (GeV) &
        &
$n^3S_1$ & all \\
\hline
\1{10}{|c|}{POSITRONIUM} \\									
\hline
EM & \1{2}{c|}{EM} &
$\ee$ & 
Ortho- & $5\times 10^{-15}$ & 0.001 & 
$\sim0.0$ &
2 & 8 \\
\hline
\1{10}{|c|}{QUARKONIUM} \\									
\hline
Strong & \1{2}{c|}{Strong} &
$u\bar u$, $d\bar d$ &
$\rho$ & 150.00 & 0.8 & 
$\sim1.0$ & 0 & 0 \\
\cline{4-10}
 & \1{2}{|c|}{\quad} &
$s\bar s$ &
$\phi$ & 4.40 & 1.0 &
$\sim0.8$ &
``1'' & ``2'' \\
\cline{3-10}
  &  & EM &
$\cc$ & $\psi$ & 0.09 &	3.1 &
$\sim0.25$ &
2 & 8 \\
\cline{4-10}
  &  &  &
$\bb$ & $\Upsilon$ & 0.05 & 9.5 &
$\sim0.08$ & 3 & 30 \\
\cline{2-10}
 & \1{2}{c|}{weak} &
$t\bar t$ & & (3000.0) & (360.) &
$<0.01$ &
0 & 0 \\
\hline
\end{tabular}
}
\end{center}
%\vskip-0.5cm
\vspace{0.5cm}
\end{table*}

\footnotetext{$^*$Invited talk presented at the 21$^{st}$ International
Symposium On Lepton And Photon Interactions At High Energies (LP03)
11-16 August 2003, Batavia, Illinois.}

\section{Quarkonia}

Quarkonium is a bound state of a quark and its antiquark.
Some properties of light and heavy quarkonia are
compared to properties of positronium in Table~\ref{tab:onia}.
Unlike positronium, light quarkonia are highly relativistic.
They also contain mixtures of quarks of different flavor
and fall apart easily into other mesons.
Charmonium ($\cc$) was the first heavy quarkonium 
discovered and is less relativistic;
the number of long-lived states below the dissociation energy 
({\it i.e.} the threshold for decay to $\DD$ meson pairs)
equals the number of long-lived positronium states.
Bottomonium ($\bb$) is even more non-relativistic and has 
a larger number of long-lived states.
The toponium system would have been completely non-relativistic.
However, weak decays of the top quark will dominate over the strong
binding and long-lived states will not be formed.
Therefore, charmonium and bottomonium play a special role 
in probing strong interactions.

The states below open flavor threshold live long enough
for electromagnetic transitions between various excitations
to occur. The electromagnetic transitions compete with
transitions mediated by the emission of soft gluons. The latter 
materialize as light hadrons. Eventually the heavy quarks
must annihilate into two or three hard gluons.
Properties of these bound states and their decays are good
testing grounds for QCD in both the non-perturbative and perturbative
regimes. 

The first heavy quarkonium bound state above 
the $D\bar D$ or $B\bar B$ threshold
acts as a factory of heavy-light mesons. 
The heavy quarks trapped in these
mesons ultimately decay via weak interactions.
This is a good place to look for physics beyond the
standard model as
discussed by Y.~Grossman
at this Symposium\rlap.\,\cite{GrossmanLP03}
Many measurements of electroweak parameters
are obscured by strong interactions.
This provides an important motivation for
trying to understand details of strong
interaction phenomena. 
P.~Lepage pointed out that at least some of the
new interactions to be discovered are
likely to be strongly coupled, further
motivating detailed studies of 
QCD\rlap.\,\cite{LepageLP03}  

Heavy quarkonia offer two small parameters:
velocities ($v$) of constituent quarks and
the strong coupling constant ($\alpha_s$) in annihilation 
and production processes.
The expansion of the full theory in these parameters allows for
effective theories of strong interactions:
in the past - purely phenomenological potential models;
more recently - NRQCD\rlap.\,\cite{NRQCD}
Lattice QCD calculations are also easier for
heavy quarkonia than for light hadrons.

\vskip-0.2cm
\section{Hadroproduction}

Annihilation of $n^3S_{1^{--}}$  
($\psi$, $\Upsilon$) states to $\mm$  and
to a lesser extent $\ee$ 
makes it possible for experimentalists to 
fish out these states from large
backgrounds in hadroproduction 
experiments.
This is how these states were (co)discovered!
There is also some access to the production of 
the excited states by addition of a transition photon or
$\pi^+\pi^-$ pair.  
So far, except for the initial discovery, 
the hadroproduction
experiments have not played an important role
in spectroscopy or decay studies, but have made interesting
production measurements.
Heavy quarkonium is also a useful 
probe for determining the structure of the target 
(e.g. to probe for gluon content or the presence of
quark-gluon plasma).

Non-relativistic QCD (NRQCD)
is a favorite theoretical approach
used to describe the production data\rlap.\,\cite{NRQCDrev}
Various diagrams are ordered 
according to powers 
of $\alpha_s$ and $v$.
The diagrams can also be classified into
color-singlet processes, in which the heavy
$Q\bar Q$ pair is produced 
colorless and therefore can directly
form a bound state, and color-octet
processes, in which the $Q\bar Q$ pair has
color which must be emitted away
by soft gluon (i.e. long distance)
interactions.
Initial attempts to describe 
charmonium production at the Tevatron
with only color-singlet processes
(CS model)\cite{CS} failed
spectacularly, revealing the importance of
color-octet diagrams.
However, the color-evaporation model 
(CEM)\rlap,\,\cite{CEM} 
which allows color-octet processes
but neglects their ranking in $v$, 
also fits the data well. 
All theoretical approaches involve
free parameters, which are fixed
by fits to the data, diminishing
their effective differences.

Polarization of produced heavy quarkonia
should lead to a better discrimination
between NRQCD and CEM. NRQCD predicts 
an increase in polarization with higher 
$p_t$, while CEM predicts no polarization
at all. The present data do not provide
any evidence for increase in polarization 
with $p_t$, however the experimental errors
are too large to draw any firm conclusions.

There is a large
range of kinematical regimes and differential 
cross sections for charmonium production
studies at HERA. Both H1 and ZEUS
contributed a number of papers on this topic
to this Symposium\rlap.\,\cite{H1ZEUS}
NRQCD predicts that matrix elements should
be universal. However, it is difficult to
reconcile all Tevatron and HERA data with a
consistent set of NRQCD matrix elements.
It is quite possible that the charm quark  
is just not heavy enough for the NRQCD
approximation to work well.
Unfortunately, data on bottomonium
production are either non-existent or have large
experimental errors.

For a more complete review of this topic
see e.g. Kr\"{a}mer\rlap.\,\cite{NRQCDrev}

\vskip-0.2cm
\section{Clean Production Environments}

\begin{figure}[htbp]
\center
\vskip-0.4cm
%\rule{2cm}{0.2mm}\hfill \rule{2cm}{0.2mm}
%\vskip 6cm
%\rule{2cm}{0.2mm}\hfill \rule{2cm}{0.2mm} 
\psfig{figure=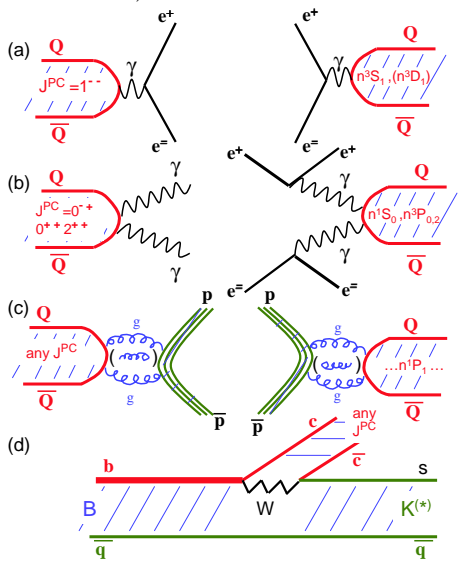,width=6.0truecm}
\caption{Clean production environments for heavy quarkonia
         together with corresponding decay modes for most of them.}
\label{fig:prod}
\vskip-0.6cm
%\vspace{-0.4cm}
\end{figure}

Most of what we know about heavy quarkonium states 
and their decays comes from experiments 
at clean production environments, 
which are time reversals of simple decay modes (see 
Fig.~\ref{fig:prod}).

Vector states ($J^{PC}=1^{--}$: $n^3S_1$, $n^3D_1$)
decay to lepton pairs and thus can be directly
formed in $\ee$ collisions. Production rates
are large compared to the other $\ee$ processes, 
thus the backgrounds are small and these
states can be studied in both inclusive and 
exclusive decay modes. Dedicated
runs are needed for each vector state. 
The other states can be
reached via photon and hadronic 
transitions, however, their scope is limited
by the transition selection rules and
branching ratios.

Spin 0 and 2 states with positive C-parity
($J^{PC}=0^{-+}, 0^{++}, 2^{++}$: $n^1S_0$, $n^3P_{0,2}$)
decay to two-photons, thus can be formed in
two-photon collisions at the  $\ee$ colliders.
No dedicated runs are needed since the two-photon flux
populates a wide range of possible quarkonia masses.
However, the flux drops quickly with the energy and
production rates for heavy quarkonia are small 
compared to the dominant $\ee$ processes.
To combat these large backgrounds the quarkonia states
must be detected in simple exclusive decay modes.
So far this formation method has succeeded only
for charmonium states.

States of any quantum numbers can decay to $p\bar p$
via two- or three-gluon annihilation, however,
such couplings are small.
Low energy $\bar p$ beams annihilating with 
proton-jet targets were successfully used to form
charmonium states. The beam energy must be tuned to form
a specific quarkonium state. Since the background
cross-sections are very large, the quarkonia must be detected
in simple exclusive final states to take advantage of 
the highly constrained kinematics.

Decays of $B$ mesons also offer a clean production
environment for charmonium states. Again states of any
quantum number can be formed.
Since the production rates are rather small,
experimentalists must restrict themselves to specific
exclusive final states to take advantage of the $B$ mass
constraint (and beam energy constraint 
if produced at the $\Y(4S)$ resonance in
$\ee$ collisions).

%\begin{figure}[htbp]

\vskip-0.2cm
\section{Results from $\bar pp$ Experiments}

A $H_2$ gas jet is used as a target for an antiproton
beam in the $\bar p$ accumulator ring.
This technique was pioneered at the CERN-ISR 
by the R704 experiment (1983-84)\rlap.\,\cite{R704}
It was later used at Fermilab 
with the E760 detector (1989-91)\cite{E760}
followed by its major upgrade E835 (1996-97, 2000).
Since this was a non-magnetic detector, it was limited to
specific final states containing electrons and photons.
Charmonium states are observed directly
as resonance peaks in the signal event yield 
measured as function of the center-of-mass energy. 
Masses and widths of the observed states can be measured
with high precision, since the accuracy depends on knowledge of
the beam energy rather than resolution of the detector.
The Fermilab experiments determined masses and widths of
the $\chi_{cJ}$ states with 
unprecedented accuracy\rlap.\,\cite{E760,E835chimw}
They also studied radiative transitions 
from these states\rlap.\,\cite{E835rad}
The most recent results include detection of the $\chi_{c0}$
resonance in decays to $\pi^0\pi^0$, interfering with 
continuum production of this final state\cite{E835pzpz}
and observation of the singlet $\eta_c(1^1S_0)$ state in annihilation
to two photons\rlap.\,\cite{E835etac}
The radial excitation of the latter was not found in scans
by E760 and E835\rlap.\,\cite{E835etacp}
 Only the region
around the mass reported by the Crystal Ball experiment\cite{CBetacp}
was scanned.

While the singlet $\eta_c(1^1S_0)$ was observed in many 
different environments,
the only hints for the singlet $h_c(1^1P_1)$ state come
from  the $\bar pp$ experiments.
R704 provided some inconclusive 
evidence for this state in 1984\rlap.\,\cite{R704hc}
Better evidence for this state, with the mass close to the
center-of-gravity mass of the $\chi_c(1^3P_J)$ states,
was reported by E760\cite{E760hc}
via the decay chain: $h_c\to \pi^0 J/\psi$, $J/\psi\to\ee$.
E835 took more scans in this mass range, with larger statistics
and an improved detector. 
Recently, rumors of absence of the $h_c$ state in the
E835 data was reported in print by several non-E835
authors\rlap.\,\cite{hcRumors} 
However, the official word from
the E835 collaboration\cite{CesterPrivate}
is that they are analyzing all available decay channels and
are not ready to make any definite statements yet.

%\begin{figure}[htbp]
\begin{figure}[htbp]
\center
%\vskip-0.7cm
Average: $(2979.9\pm1.0)$ MeV \\
CL=0.5\%\  Scale Factor=1.5 \\[-1.2truecm]
\hbox{
\psfig{figure=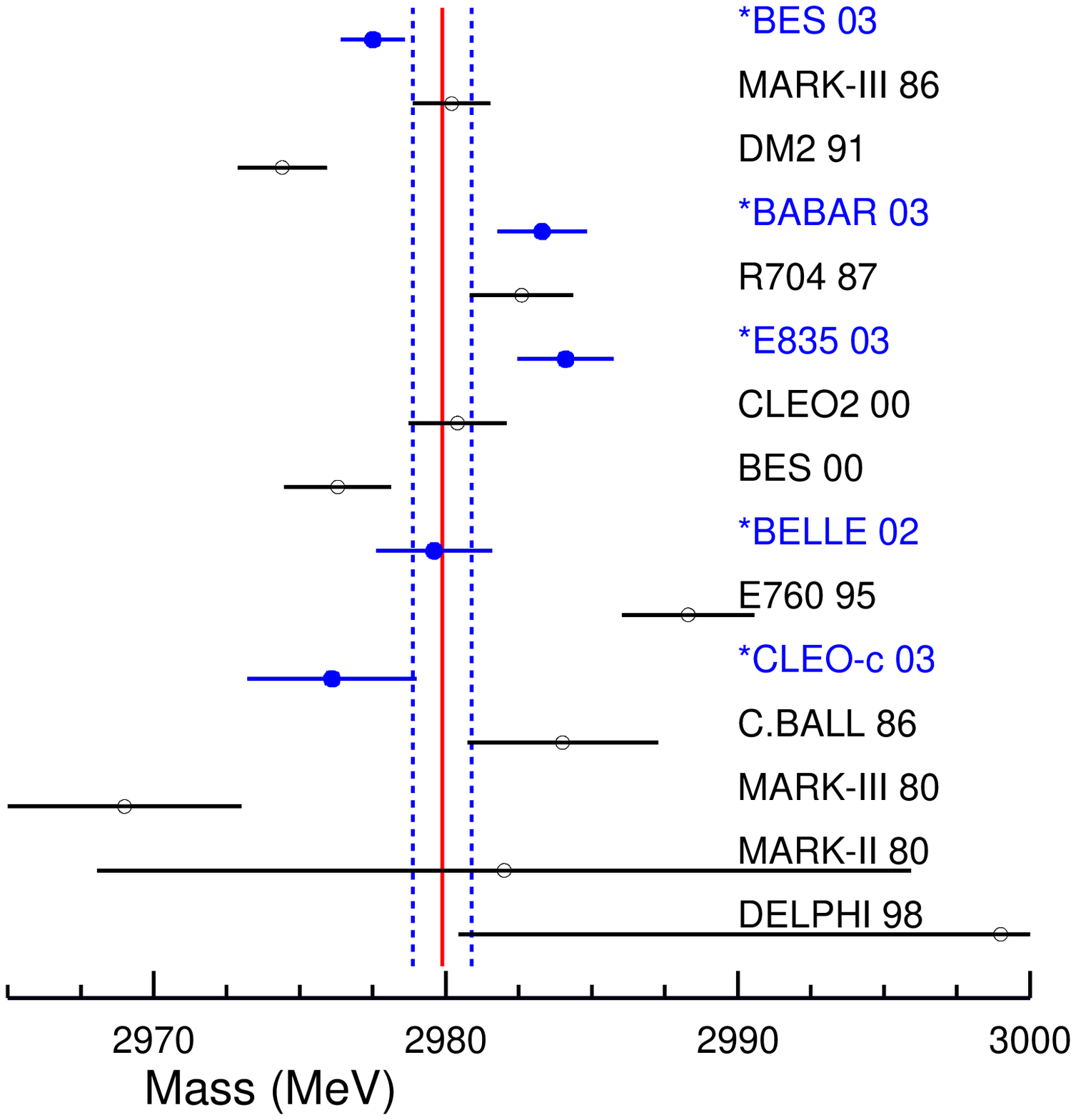,width=7.0truecm,height=7.0truecm}
\hskip-1.0truecm
\psfig{figure=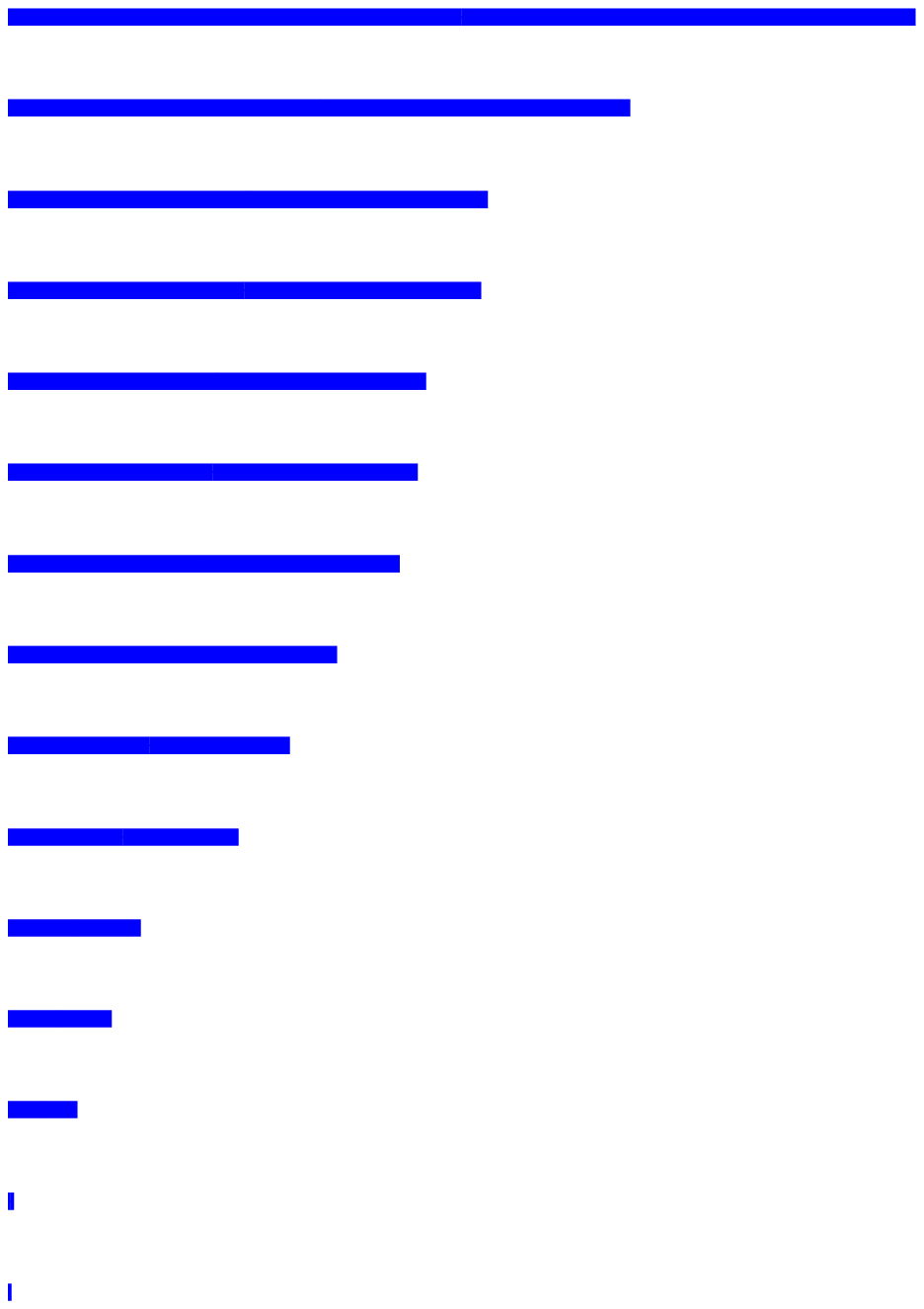,width=2.0truecm,height=7.14truecm}
}
\vskip-0.2cm
\caption[]{Measurements of the $\eta_c$ mass. 
         The new measurements 
         (BES 03\rlap,\,\cite{BESetac} BABAR 03\rlap,\,\cite{BABARetac}
          E835 03\rlap,\,\cite{E835etac} BELLE 02\cite{BELLEetac}
         and CLEO-c 03\cite{CLEOcetac})
         are indicated by a star.
         See PDG 2002\cite{PDG2002}
          for the references to the
         older measurements displayed here.
         The vertical bars indicate the 
         average value (solid) and its error
         (dashed). The error includes the
         scale factor.
         The thick horizontal bars on the right
         give the relative weight of each experiment
         into the average value.}
\vskip-0.6cm
%\vspace{-0.5cm}
\label{fig:etacmass}
\end{figure}

\vskip-0.2cm
\section{Charmonium Singlet States}

\begin{figure}[htbp]
\center
\vskip-0.2cm
\psfig{figure=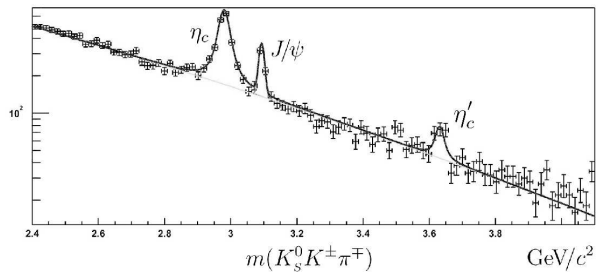,width=8.0truecm}
\vskip-0.25cm
\caption[]{Observation of $\gamma\gamma\to\eta_c(1S)$
         and $\gamma\gamma\to\eta_c(2S)$ 
         by the BABAR experiment\rlap.\,\cite{BABARetac}         
         A peak due to $\ee\to\gamma J/\psi$ 
         is also visible. These data serve as a
         precision determination of the $\eta_c(1S)$
         width and confirmation of the $\eta_c(2S)$
         state at its new mass value.}
\label{fig:babar-etacp}
\vskip-0.6cm
\end{figure}

\begin{figure}[htbp]
\center
%\vskip-0.3cm
\center
Average: $(25.0\pm3.3)$ MeV \\
CL=0.05\%\  Scale Factor=1.8 \\[-1.2truecm]
\hbox{
\psfig{figure=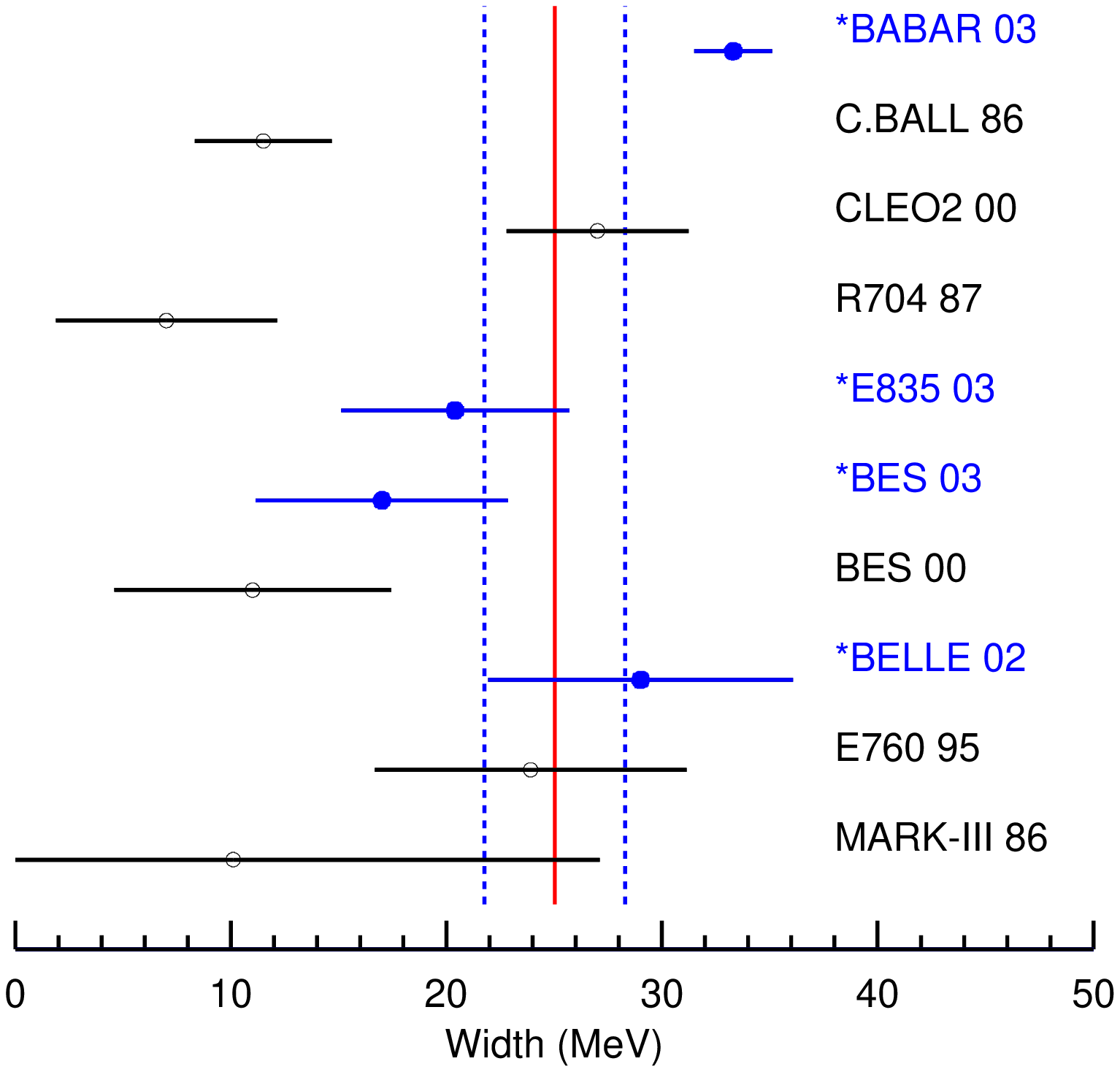,width=7.0truecm,height=7.0truecm}
\hskip-1.0truecm
\psfig{figure=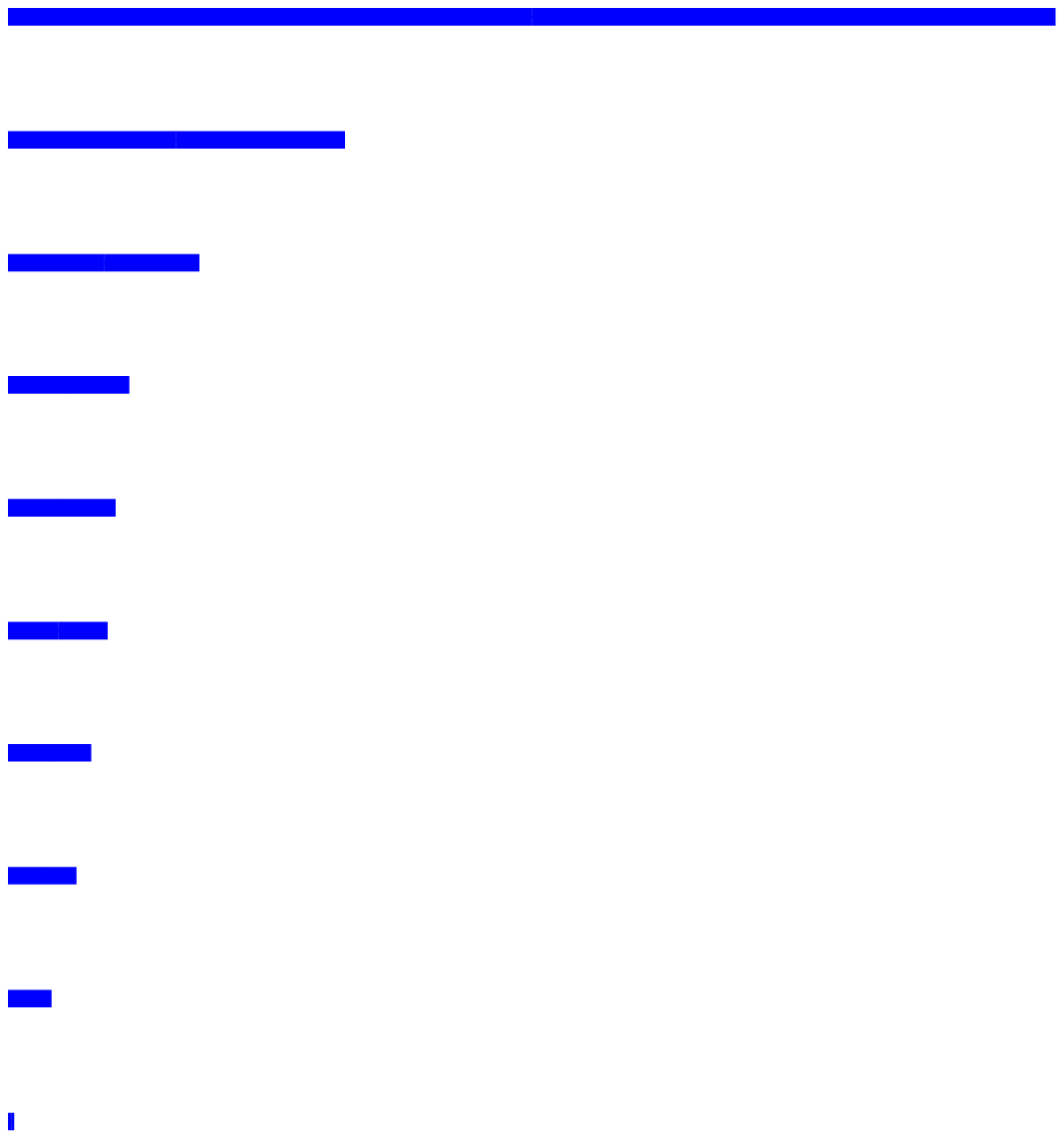,width=2.0truecm,height=7.14truecm}
}
\vskip-0.2cm
\caption{Measurements of the $\eta_c$ width. 
         See Fig.~\protect\ref{fig:etacmass} 
         for references and explanation.}
\label{fig:etacwidth}
\vskip-0.6cm
\end{figure}

There are many new measurements related
to the charmonium singlet states, in addition to
the results from the $\bar pp$ experiments
mentioned in the previous section.
For example, there are 5 new measurements
of the  $\eta_c(1^1S_0)$ 
mass\cite{E835etac,BESetac}$^-$\cite{CLEOcetac} 
(see Fig.~\ref{fig:etacmass}).
BES-II\rlap,\,\cite{BESetac}
which reconstructed the
$\eta_c$ in 5 different decay modes
in the $J/\psi$ data, achieved the smallest
overall error. 
Consistency of the old 
and new measurements is questionable.
A new preliminary measurement 
of the $\eta_c$ width
by BABAR\cite{BABARetac}
constitutes an even
bigger experimental puzzle
(see Figs.~\ref{fig:babar-etacp}-\ref{fig:etacwidth}).
The measurement error claimed by BABAR
is much smaller than in any
other measurement. At the same time the
central value, $(33.3\pm2.5\pm0.5)$ MeV, 
is the highest ever measured;
completely inconsistent with the previous
world average value\rlap,\,\cite{PDG2002}
$(16.0^{+3.6}_{-3.2})$ MeV.

There have also been dramatic developments
concerning the $\eta_c(2^1S_0)$ state.
The Crystal Ball experiment at SPEAR 
claimed observation of this state over 20 years ago\rlap.\,\cite{CBetacp}
They observed a peak in the inclusive photon energy spectrum
from $\psi(2^3S_1)$ decays, which they interpreted as a
magnetic dipole photon transition, 
$\psi(2S)\to\gamma\eta_c(2S)$\rlap.\,\cite{CBetacp}
The Crystal Ball result corresponded to the hyperfine
mass splitting in the radial excitation, which was
only slightly smaller than in the ground
state (92 MeV vs.\ 117 MeV).
Last year BELLE observed both $\eta_c$ states
in $B\to K\eta_c(nS)$, 
$\eta_c(nS)\to K_s K^+\pi^-$\rlap.\,\cite{BELLEetacp}
The $\eta_c(2S)$ state appeared at much higher
mass than in the Crystal Ball measurement,
thus reducing the corresponding hyperfine splitting.
BELLE also observed both $\eta_c$ states 
in the spectrum of the mass recoiling against 
$J/\psi$ in continuum $\ee$ annihilation to 
$J/\psi X$\rlap.\,\cite{BELLEdoubleccOld} 
These results have been updated this
year with larger statistics\rlap.\,\cite{BELLEdoubleccNew}
The $\eta_c(2S)$ mass obtained from these data
differs by 1.9 standard deviations from
the other BELLE result, but is still significantly
higher than the Crystal Ball value.

The production rate for charmonium pairs observed
by BELLE in the latter analysis 
is surprisingly high. They also observe
a large $\ee$ annihilation rate to $J/\psi$ plus
open charm. They claim 
$(82\pm15\pm14)\%$ of prompt $J/\psi$ production
($P_{J/\psi}>2$ GeV) at 10.6 GeV center-of-mass
energy is associated with another $c\bar c$ pair
in the event\rlap.\,\cite{BELLEdoubleccNew} 
This rate is an order of magnitude
higher than that predicted theoretically, and 
constitutes the biggest puzzle in 
quarkonium production physics\rlap.\,\cite{doubleccPuzzle}

This year the BABAR and CLEO experiments confirmed  
the $\eta_c(2S)$ at its heavier mass by formation
in the two-photon collision
process\cite{BABARetac,CLEOetacp}
(see Fig.~\ref{fig:babar-etacp}).
The discrepancy with the Crystal Ball mass measurement
is resolved by the CLEO-c experiment, which
has been able to remeasure the inclusive photon spectrum
from the $\psi(2S)$ for the first time since
the Crystal Ball experiment (see the next section). 
While CLEO-c agrees well with the
Crystal Ball on all other E1 and M1 transitions
from the $\psi(2S)$, the direct M1 transition to 
the $\eta_c(2S)$ is not observed at the photon
energy claimed by the Crystal Ball.
The upper limit on the rate for this transition
set by CLEO-c disagrees with the rate measured by
the Crystal Ball. Therefore, the peak observed
by Crystal Ball could not be due to
$\eta_c(2S)$ production.

\begin{figure}[htbp]
%\vskip-0.3cm
\center
Average: $(3637.7\pm4.4)$ MeV \\
CL=14\%\  Scale Factor=1.3 \\[-1.2truecm]
\hbox{\quad
\hskip-1.0truecm
\psfig{figure=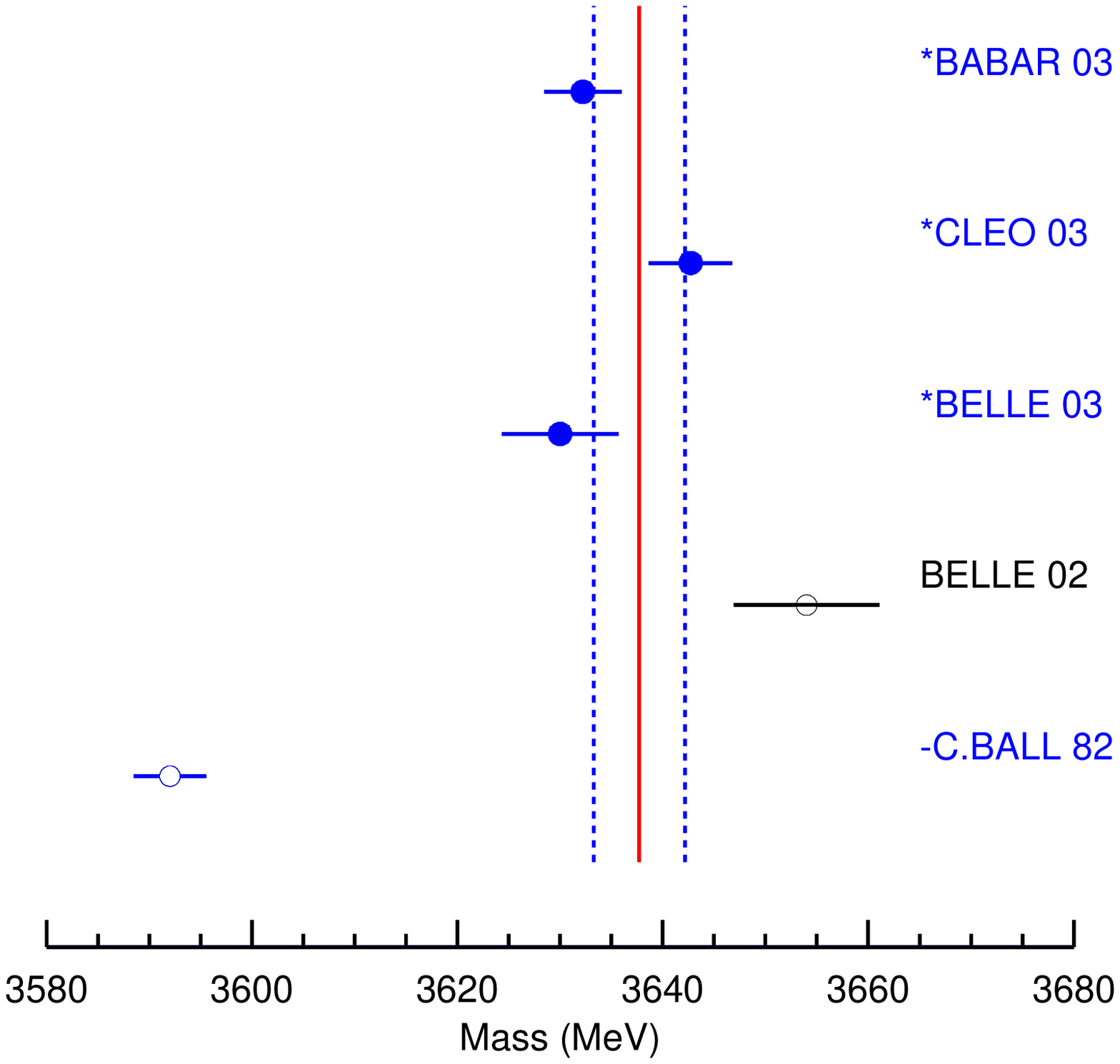,width=7.0truecm,height=7.0truecm}
\hskip-0.7truecm
\psfig{figure=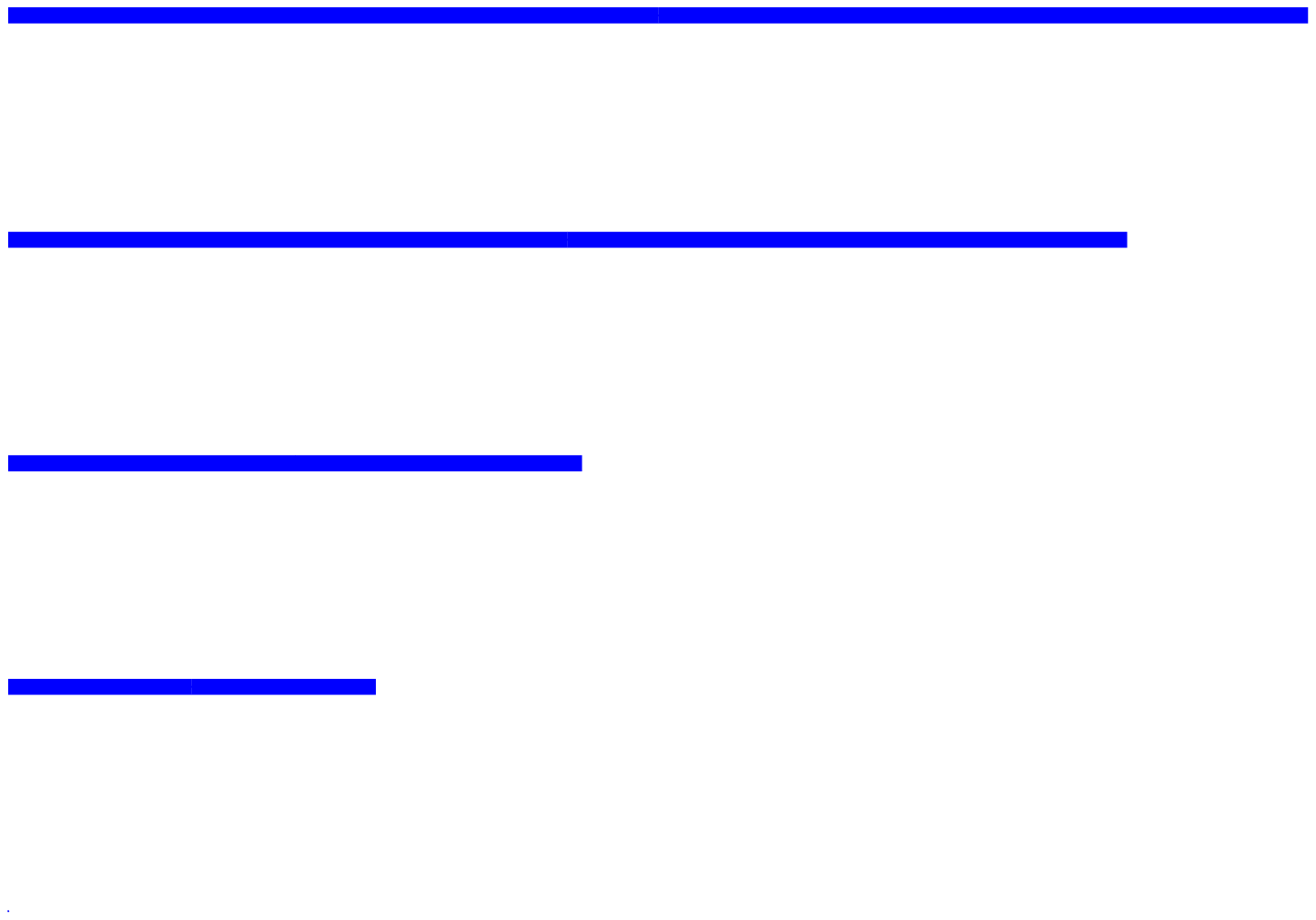,width=2.0truecm,height=7.3truecm}
}
\vskip-0.2cm
\caption{Measurements of the $\eta_c(2S)$ mass
         (see the text for the references).
         The thick horizontal bars on the right
         give the relative weight of each experiment
         into the average value.
         The Crystal Ball measurement was excluded from 
         the average.}
\label{fig:etacpmass}
\vskip-0.6cm
\end{figure}

Measurements of the $\eta_c(2S)$ mass
are summarized graphically 
in Fig.~\ref{fig:etacpmass}. 
All new mass
measurements are consistent with each other.
The total width of this state, based on the
BELLE\cite{BELLEetacp}
and BABAR\cite{BABARetac} results, is $(19\pm10)$ MeV.

\begin{figure}[htbp]
\center
\vskip-1.2cm
\psfig{figure=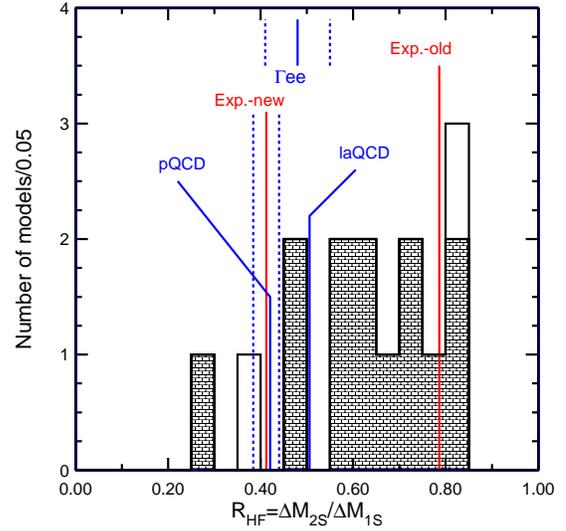,width=8.0truecm}
\vskip-0.2cm
\caption[]{Measurements and predictions for
         the ratio of the hyperfine mass splittings
         in the $2S$ and $1S$ states in charmonium.
         The old and new experimental values
         are indicated by vertical bars. The dashed
         lines around the new value show the experimental
         error. The histogram shows predictions of 
         a sample of potential models\rlap.\,\cite{hyperfineRatioPlot} 
         Only models
         which gave a good fit to the masses of all known
         charmonium states were included. 
         The shaded part of the histogram represents 
         the models published before the new measurements of
         the $\eta_c(2S)$ mass. 
         Values predicted by scaling from the measured 
         leptonic widths ratio\cite{BadalianBakker} ($\Gamma_{ee}$),
         by perturbative QCD alone\cite{RecksiegelSumino}
         (pQCD) and by lattice QCD\cite{LatticeHF} (laQCD)
         are also indicated.}
\vskip-0.6cm
\label{fig:hyperfine}
\end{figure}

\def\RHF{R_{HF}}
The new mass of the $\eta_c(2S)$ state 
decreases the hyperfine mass splitting
for the $2S$-states by a factor of 2
compared to the old value.
Since a wrong value 
prevailed for 20 years, it is interesting to
check for an experimental bias on the
phenomenological predictions for this
splitting. A sample of potential model
predictions for the ratio of the
hyperfine mass splitting, $\RHF\equiv$
$(M_{\psi(2S)}-M_{\eta_c(2S)})/
 (M_{\psi(1S)}-M_{\eta_c(1S)})$,
is compared to the old ($R=0.79$) and
the new ($\RHF=0.412\pm0.028$) experimental
values in Fig.~\ref{fig:hyperfine}. 
It appears that many old calculations
were stretched to accommodate the old 
result, though there were a few that had the
courage to contradict the data.
The ratio of hyperfine mass splitting
can be related to the ratio of the leptonic
widths  of the triplet 
states ($\Gamma_{ee}$) using perturbative QCD. 
The new value agrees well with
the prediction by 
Badalian and Bakker\rlap,\,\cite{BadalianBakker}
$\RHF=(0.48\pm0.07)$, 
based on this approach.
Instead of using the measured values
of $\Gamma_{ee}$,
Recksiegel and Sumino\cite{RecksiegelSumino}
extended the use of 
perturbative QCD to the extraction of the
interquark potential at short distances
relevant for the spin-spin forces.  
Their prediction for
the hyperfine mass splitting ratio,
$\RHF=0.42$, is in good agreement with the data,
but the absolute values for the mass 
splittings are underestimated.
Lattice QCD calculations\cite{LatticeHF}
predicted $\RHF=0.5$, not 
far from the measured value.

\vskip-0.2cm
\section{Radiative Transitions}

\begin{figure}[htbp]
\center
\vskip-0.6cm
\psfig{figure=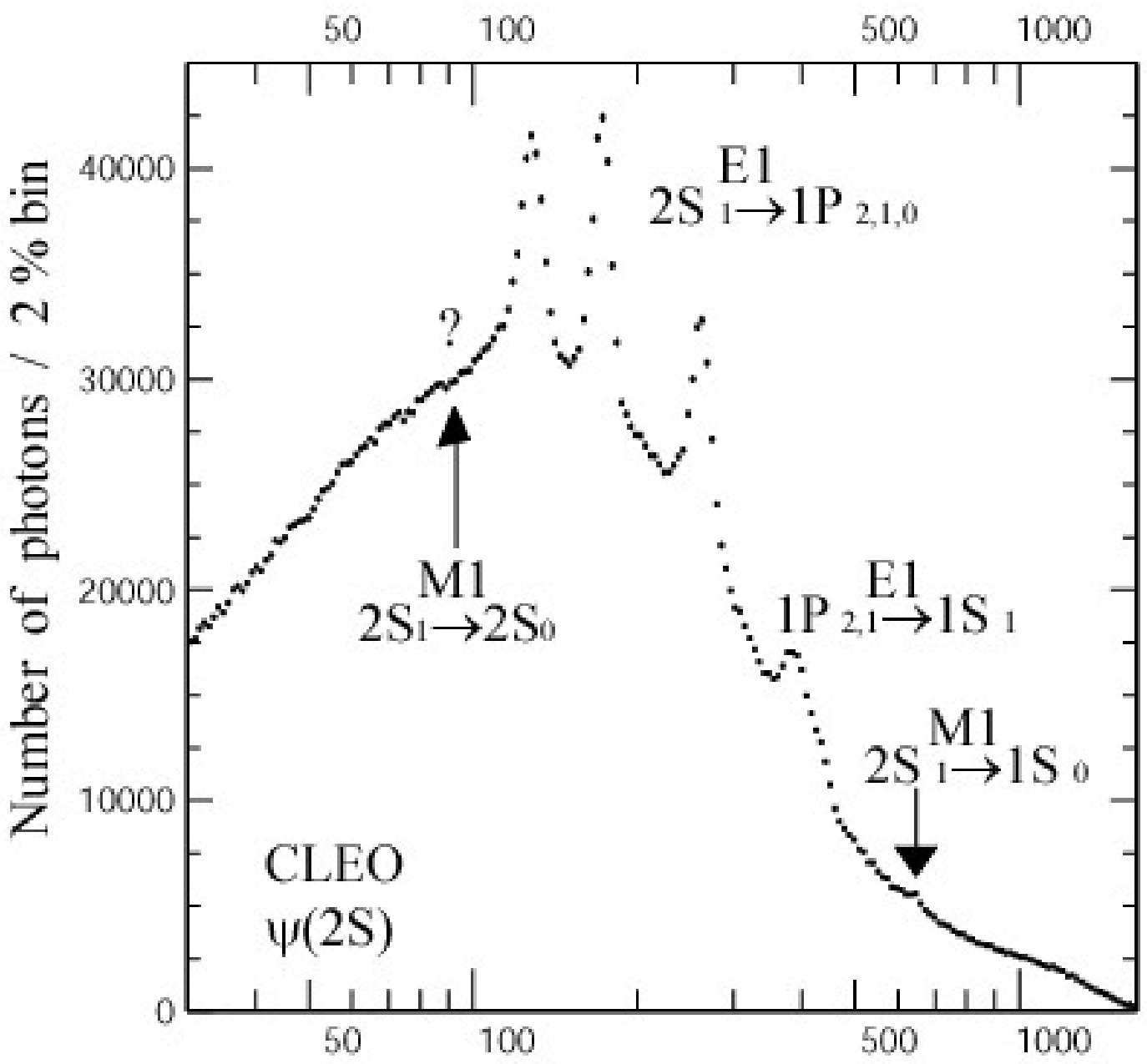,width=8.0truecm,height=5.5truecm}
\vskip-0.1cm
\psfig{figure=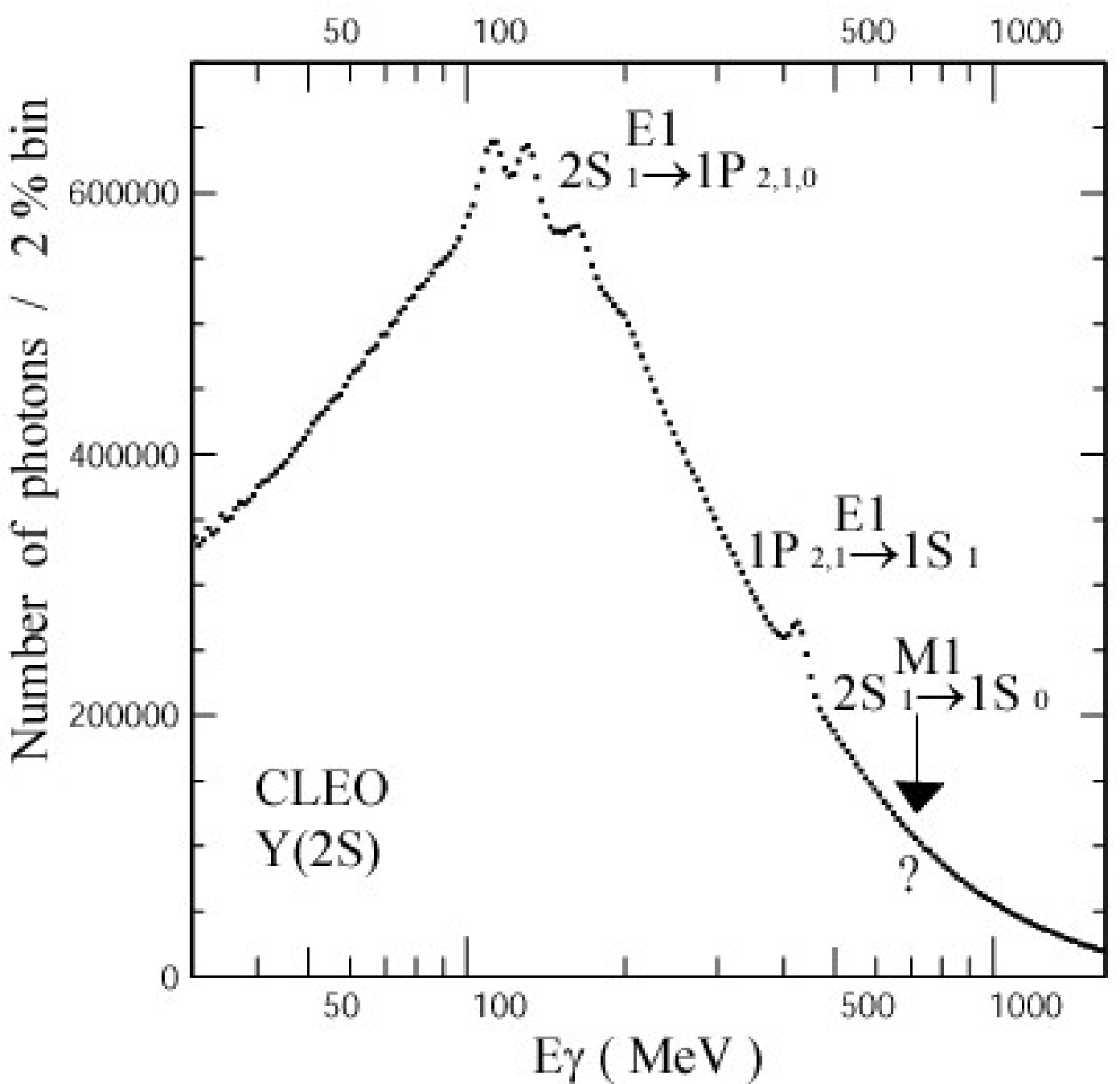,width=8.0truecm,height=5.5truecm}
\vskip-0.2cm
\caption{Inclusive photon spectrum 
         from $2^3S_1$ decays
         in the $\cc$ (top) and $\bb$ (bottom) systems 
         measured with the CLEO detector.
         The data correspond to about 1.5M $\psi(2S)$ 
         and 9M $\Upsilon(2S)$ decays.}
\vskip-0.7cm
%\vspace{-0.4cm}
\label{fig:inclusivephotons}
\end{figure}

Since sizes of heavy quarkonia are small or comparable to the
wavelengths of transition photons, selection rules are given
by multipole expansion. As the system is non-relativistic, electric
dipole transitions ($\Delta L\!=\!1$, $\Delta S\!=\!0$) 
are much stronger than magnetic 
dipole transitions ($\Delta L\!=\!0$, $\Delta S\!=\!1$).
The latest improvements in measurements of these transitions
come mostly from the CLEO experiment, which is equipped with
an excellent CsI(Tl) calorimeter. Recently, after increasing 
statistics of the $\Upsilon(1,2,3S)$ data samples by 
an order of magnitude, CLEO turned the beam energy down and 
collected also $\psi(2S)$ data. This was the beginning of 
the CLEO-c phase.  
Inclusive photon spectra collected at $\psi(2S)$ and $\Upsilon(2S)$
by CLEO are compared in Fig.~\ref{fig:inclusivephotons}.
The dominant peaks are due to E1 transitions, $2^3S_1\to\gamma 1^3P_J$.
The peaks in bottomonium system appear smaller because of the 
increased $\pi^0\to\gamma\gamma$ background induced by higher
particle multiplicities. The peaks are more crowded together 
reflecting the smaller fine structure of the $1^3P_J$ states in the more
non-relativistic $\Upsilon$ system.
The E1 cascade lines, $1^3P_J\to\gamma 1^3S_1$ are also visible.
The charmonium data also reveal a small peak due to
the hindered M1 transition, $2^3S_1\to\gamma1^1S_0$.
This is the first confirmation of this transition since the original
observation by the Crystal Ball. As discussed in the previous
section, the direct M1 transition, $2^3S_1\to\gamma2^1S_0$, is ruled out
for the photon energy, $\eta_c(2S)$ width 
and rate claimed by the Crystal Ball.
The preliminary CLEO-c results\cite{CLEOinternal}
for the $\psi(2S)$ photon lines are: 
$\B(\psi(2^3S_1)\to\gamma\chi_c(1^3P_J))=
 \{9.75\pm0.14\pm1.17,9.64\pm0.11\pm0.69,9.83\pm0.13\pm0.87\}\%$
for $J=\{2,1,0\}$,
$\B(\psi(2^3S_1)\to\gamma\eta_c(1^1S_0))=
   (0.278\pm0.033\pm0.049)\%$
and 
$\B(\psi(2^3S_1)\to\gamma\eta_c(2^1S_0))<0.2\%$ for
$E_\gamma=(91\pm5)$ MeV and $\Gamma(\eta_c(2S))=8$ MeV (90\%\ CL).

\begin{figure}[htbp]
\center
\vskip-0.1cm
\psfig{figure=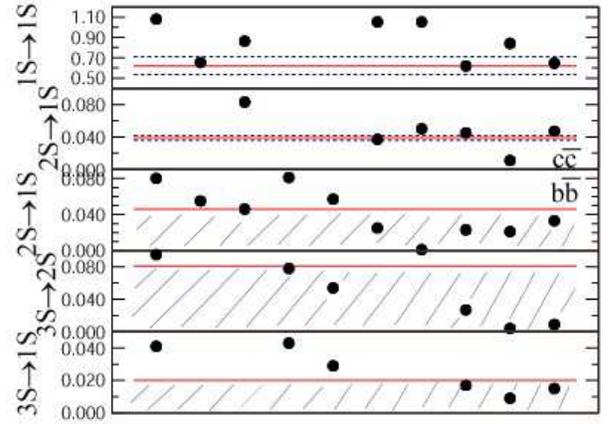,width=8.0truecm}
\vskip-0.05cm
\caption[]{Measured and predicted values of matrix elements
         for M1 transitions in heavy quarkonia.
         The measured values are calculated
         from the world average branching ratios and 
         total widths of the triplet $S$ states.
         The central value and error bars for the $\cc$ data
         are indicated with solid and dashed
         lines respectively. For the $\bb$ data, 
         allowed ranges from the preliminary CLEO
         analysis\cite{CLEOinternal} are shaded (90\%\ CL).
         The theoretical predictions\cite{m1preds} (points) 
         are ordered according to the publication date.}
%\vspace{-0.4cm}
\label{fig:m1}
\end{figure}

\begin{figure}[htbp]
\center
\vskip-1.3cm
\psfig{figure=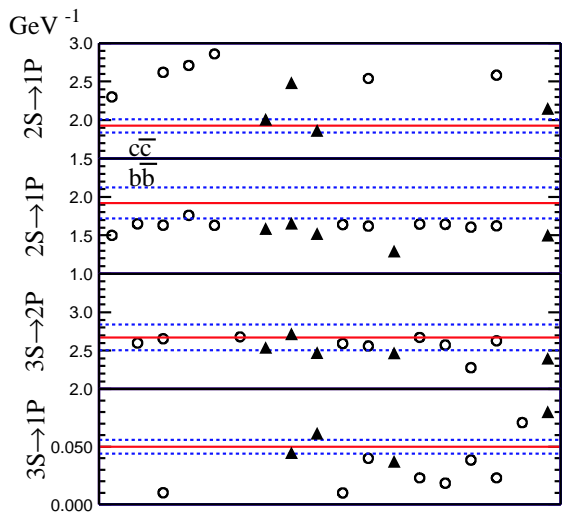,width=8.0truecm}
\vskip-0.6cm
\caption[]{Measured and predicted values of matrix elements
         for E1 transitions in heavy quarkonia averaged over
         different spins of the triplet $P$ states. 
         The measured values are calculated
         from the world average radiative branching ratios and 
         total widths of the triplet $S$ states.
         The central values and error bars for the measured values
         are indicated with solid and dashed
         lines respectively. Circles (triangles) show non-relativistic 
         (relativistic) calculations. The relativistic calculations are 
         averaged over spins with the same weights as the data.
         The predictions\cite{e1preds} 
         are ordered according to the publication date.
         }
\vskip-0.6cm
\label{fig:e1}
\end{figure}

None of M1 transitions, hindered or direct, are observed
for the bottomonium. This is not a big surprise, since
the backgrounds are much higher and the expected branching
ratios are smaller in the more non-relativistic $\bb$ system.
Searches for the singlet $\eta_b$ state in two-photon collisions
at LEP has also yielded upper limits only\rlap,\,\cite{LEPetab}
thus no singlet states
have been observed in bottomonium so far.
The M1 rate measurements can be translated into values of
the corresponding matrix elements, which are compared to 
theoretical predictions in Fig.~\ref{fig:m1}.
The matrix elements for the hindered M1 transitions are expected
to be very small, since they are generated by relativistic and
finite size corrections.
Therefore, they are difficult to predict. 
Only very recent calculations are consistent with all charmonium and
bottomonium data.
Similar comparisons for the E1 matrix elements is shown
in Fig.~\ref{fig:e1}. 
Non-relativistic calculations (hollow circles) overestimate
the E1 rates in charmonium. The predictions with
relativistic corrections (filled triangles) can reproduce the data.
Relativistic effects in the dominant E1 transitions in bottomonium
are much smaller. However, relativistic calculations are needed
to reproduce the rare $3^3S_1\to\gamma1^3P_J$ transition rate.
This matrix element is small because of the cancellations
in the integral for the E1 operator between the initial and the
final state wave functions\cite{EichtenGottfried77}
for which the number of nodes differ by two.

%exclusives ?
%give refernces in caption only

\vskip-0.2cm
\section{Hadronic Transitions}

Heavy quarkonia can also change excitation levels  by
emission of soft gluons, that turn into light hadrons.
The multipole expansion approach has proven to be useful also for
hadronic transitions\cite{Gottfried,Yan80}
explaining their gross characteristics. 
The most prominent are $\pi\pi$ transitions among
$n^3S_1$ states, which can be mediated by two-gluon emission of 
the E1$\cdot$E1 type. 
In fact, these are dominant decays for both $\psi(2S)$ and $\Upsilon(2S)$.
The ratio of the measured widths for the these transitions
agrees with a suppression by about a factor of 10 
predicted by the multipole expansion model\cite{Yan80}
due to the smaller size of the $\bb$ system. 
The multipole expansion model is also able to explain
the $M(\pi\pi)$ distributions for these transitions, which 
follow the same pattern and peak at high values.
Dipion transitions from $\Upsilon(3S)$ to $\Upsilon(2S)$ and
$\Upsilon(1S)$ are observed too. They have smaller rates 
either due to the phase space suppressions or 
cancellations in the dipole matrix element integral for the 
$3S\to\pi\pi 1S$ transition.
The $M(\pi\pi)$ distribution peaks at low and high mass values for 
the latter transition,
which reveals some dynamics beyond the multipole expansion approach.
%  S and D wave discussion?
In charmonium, an
$\eta$  transition
has been observed between the triplet $2S$ and $1S$ states\rlap.\,\cite{PDG2002}
This transition is of a magnetic type (E1$\cdot$M2),
thus it has a smaller rate than the $\pi\pi$ transition.
A $\pi^0$ transition is also observed, but with a tiny
rate due to isospin violation.
None of these transitions are 
observed between the $\Upsilon$ states\rlap,\,\cite{CLEOinternal}
which is not surprising since there is an additional suppression 
by at least a factor $1/m_Q^2$ for this type 
of transition.

\begin{figure}[htbp]
\center
\vskip-1.2cm
\psfig{figure=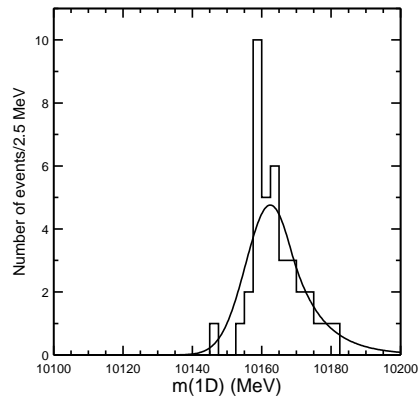,width=6.0truecm}
\vskip-0.2cm
\caption{Recoil mass against the lowest energy
         photons in four-photon cascade between
         the $\Y(3S)$ and $\Y(1S)$, consistent
         with $\chi_b(2P_{2,1})$ and $\chi_b(1P_{2,1})$
         as intermediate states, as observed by CLEO.
         The cascades via
         the $\Y(2S)$ state are suppressed, thus
         the remaining events are $\Y(1D)$ candidates.
         The recoil mass is a measure of the $\Y(1D)$ mass.
         The fit to the data (solid line) assumes production
         of just one  $\Y(1D)$ state, with a natural width
         much smaller than the experimental resolution.
         The observed events are most likely dominated by the
         production of the $\Y(1D_2)$ state. The preliminary
         CLEO results give $(10161.1\pm0.6\pm1.6)$ MeV
         for the mass of this state.}
\label{fig:m1d}
\vskip-0.6cm
%\vspace{-0.4cm}
\end{figure}

The transitions among triplet $nS$ states 
discussed above were first observed over 20 years
ago. 
A large number of other hadronic 
transitions are possible\rlap,\,\cite{EichtenGottfried77}
especially for the $\bb$ system which has a large number of long-lived
states. 
Such transitions had not been observed until recently,
when CLEO reported the first observation of 
$\chi_b(2^3P_{2,1})\to\omega\Upsilon(1S)$ 
transitions\rlap.\,\cite{chiomega}
The phase-space for these transitions is so small,
that this decay is forbidden for the $\chi_b(2^3P_0)$.
The measured branching ratios for 
the $\chi_b(2^3P_{1})$ and $\chi_b(2^3P_{2})$ states 
are of the order of a couple of percent 
($(1.6\pm0.3\pm0.2)\%$ and $(1.1\pm0.3\pm0.1)\%$ respectively),
in spite of the phase space suppression, 
which reveals that the underlying transition is 
quite strong. In fact, this transition is of
chromoelectric type, E1$\cdot$E1$\cdot$E1
(three gluons are needed to generate it).
Voloshin\cite{VoloshinOmega}
pointed out that since the matrix element does not
depend on the spin of the $2P_J$ state, transition branching
ratios for $J=1$ and $J=2$ should be comparable (as the phase
space factors approximately cancel the effect due to the smaller
total width of the $J=1$ state).
The data are consistent with this prediction.

As hadronization probabilities are difficult to estimate,
uncertainties in absolute rate predictions for
hadronic transitions are usually very large. 
When resonances dominate the transition, there are often
no theoretical predictions for the rate.
For example, there are no rate estimates for the
$\omega$ transition discussed above. 
Predictions for other yet unobserved types of 
transitions vary by orders of magnitude.
The predictions based on a model
developed by Yan\cite{Yan80} and
Kuang\cite{YanKuang81} tend to be much
larger than the predictions 
based on a different approach introduced
by Voloshin, Zakharov, Novikov and Shifman\rlap.\,\cite{NSZV}

\begin{figure}[htbp]
\center
\vskip-1.2cm
\psfig{figure=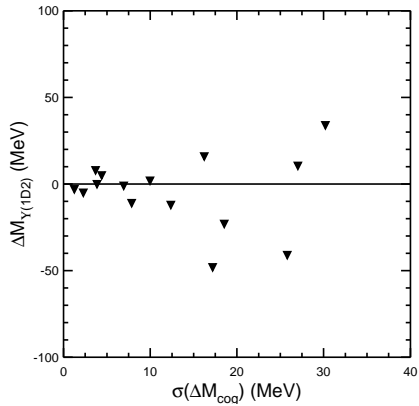,width=6.0truecm}
\vskip-0.2cm
\caption[]{The difference between the predicted and the measured
 $\Upsilon(1^3D_2)$ mass (vertical axis) versus overall quality
 of the potential model
 expressed as the r.m.s. of the differences for
 the center-of-gravity masses of long-lived 
 $\bb$ states (horizontal axis).
 Predictions for the 1S, 2S, 3S, 1P, 2P and 1D masses are included
 in the r.m.s. calculation 
 for all models displayed here\rlap.\,\cite{y1d2pot}}
\label{fig:bbdmass}
\vskip-0.6cm
\end{figure}

\begin{figure}[htbp]
\center
\vskip-1.2cm
\psfig{figure=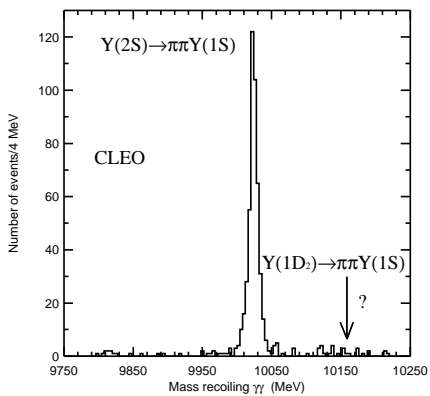,width=6.0truecm}
\vskip-0.2cm
\caption{Search for $\Upsilon(1^3D_2)\to\pi^+\pi^-\Upsilon(1S)$ transitions.
         Recoil mass against two photons in $\gamma\gamma\pi^+\pi^-l^+l^-$
         events is plotted. 
         The expected signal position is indicated by the arrow.
         }
\label{fig:y1dpipi}
\vskip-0.6cm
\end{figure}

One suitable place to test various predictions
are dipion transitions from $1^3D_J$
to $1^3S_1$.
Last year
the $\Upsilon(1^3D_J)$ states were 
discovered by CLEO 
in the four-photon cascade:\cite{Y1DAmsterdam}
$\Upsilon(3^3S_1)\to\gamma\chi_b(2^3P_{\JTwo})$,
$\chi_b(2^3P_{\JTwo})\to\gamma\Upsilon(1^3D_{\JD})$,
$\Upsilon(1^3D_{\JD})\to\gamma\chi_b(1^3P_{\JOne})$,
$\chi_b(1^3P_{\JOne})\to\gamma\Upsilon(1^3S_1)$
followed by $\Upsilon(1^3S_1)\to l^+l^-$.
CLEO has analyzed more data since then.
The recoil mass distribution against the first
two photons in the cascade is consistent with 
the detector resolution (Fig.~\ref{fig:m1d}).
Thus the data are dominated by production of 
just one $\Upsilon(1^3D_{\JD})$ state.
Theoretically, production of the $\JD=2$ 
state is expected to dominate. This spin assignment is
also favored by the experimental data\rlap.\,\cite{Y1DAmsterdam}
The measured mass
fits the predictions of the potential models
well, especially those which provide a good
fit to the other narrow $\bb$ states.
This is illustrated in Fig.~\ref{fig:bbdmass}.
The lattice QCD calculations also reproduce the
observed mass very well\rlap.\,\cite{LepageLP03}
CLEO determines\cite{CLEOinternal} 
the four-photon cascade 
branching ratio to be
$(2.6\pm0.5\pm0.5)\times 10^{-5}$.
Within the errors this rate is
consistent with the predictions by
Godfrey and Rosner\rlap,\,\cite{GodfreyRosner}
$3.8\times 10^{-5}$ ($2.6\times 10^{-5}$ for
$\JD\!=\!2$ alone),
based on the E1 matrix elements
and estimates of hadronic widths of
the $\Upsilon(1^3D_{\JD})$ and 
$\chi_b(2^3P_{\JTwo})$ states.
CLEO also looked for 
$\Upsilon(1^3D_{\JD})\to\pi^+\pi^-\Upsilon(1^3S_1)$
replacing the third and fourth photons in 
the cascade with a $\pi^+\pi^-$ pair.
The recoil mass against the remaining two photons
in the cascade is plotted 
in Fig.~\ref{fig:y1dpipi}. 
Since no signal is found, upper limits
on the product of branching ratios are set.
They are compared in Table~\ref{tab:y1dpipi}
to the theoretical
predictions calculated by
Rosner\cite{RosnerY1Dpipi}
using predictions for  
$\Upsilon(1^3D_{\JD})\to\pi^+\pi^-\Upsilon(1^3S_1)$
width by various authors\cite{YanKuang81,Moxhay88,Ko93}
together with the E1-photon
matrix elements
and estimates of hadronic widths of
the $\Upsilon(1^3D_{\JD})$ and 
$\chi_b(2^3P_{\JTwo})$ states.
Large 
$\Upsilon(1^3D_{\JD})\to\pi^+\pi^-\Upsilon(1^3S_1)$
widths, as predicted by Kuang-Yan\rlap,\,\cite{YanKuang81} are
ruled out. 
Better experimental sensitivity is needed to
test the other models. 
Voloshin pointed out\cite{VoloshinEta}
that 
$\Upsilon(1^3D_{\JD})\to\eta\Upsilon(1^3S_1)$
transition may be of comparable strength to the
$\pi\pi$ transition between these states.
CLEO doesn't find this transition either and
sets a preliminary upper limit of
$\B(\Upsilon(3^3S_1)\to\gamma\chi_b(2^3P_{\JTwo}))$ 
$\B(\chi_b(2^3P_{\JTwo})\to\gamma\Upsilon(1^3D_{\JD}))$
$\B(\Upsilon(1^3D_{\JD})\to\eta\Upsilon(1^3S_1))$
$<2.3\times 10^{-4}$ at 90 \%\ CL.

\begin{table}
%\vskip-0.6cm
\caption[]{Theoretical predictions and preliminary 
CLEO upper limits (at 90\%\ CL)\ \ on \ \
$\B(\Upsilon(3^3S_1)\to\gamma\chi_b(2^3P_{\JTwo}))$
$\B(\chi_b\\
(2^3P_{\JTwo})\to\gamma\Upsilon(1^3D_{\JD}))$
$\B(\Upsilon(1^3D_{\JD})\to\pi^+\pi^-\Upsilon(1^3S_1))$
in units of $10^{-4}$.
The theoretical predictions come from 
the paper by Rosner\rlap.\,\cite{RosnerY1Dpipi}
The first row corresponds to the cascade via the
$\JD=2$ state observed in the four-photon cascade by CLEO.
The second row corresponds to any $\Upsilon(1^3D_{\JD})$
state with mass in the $10140-10180$ MeV interval.}
\label{tab:y1dpipi}
%\vspace{-0.5cm}
\vspace{-0.3cm}
\begin{center}
\def\1#1#2#3{\multicolumn{#1}{#2}{#3}}
\tablefont{
\begin{tabular}{|l|r|r|r|r|}
\hline
   &    CLEO       & \1{3}{c|}{$\Gamma_{\pi\pi}$ model} \\
\cline{3-5}
   &               & Kuang-Yan\cite{YanKuang81} 
                   & Moxhay\cite{Moxhay88}
                   & Ko\cite{Ko93} \\ 
\hline
$\Upsilon(1^3D_{2})$   & $<1.1$ &  9.2 & 0.049 & 0.39 \\
$\Upsilon(1^3D_{\JD})$  & $<2.7$ & 17.7 & 0.094 & 0.75 \\
\hline
\end{tabular}
}
\end{center}
\vskip-0.6cm
%\vspace{-0.3cm}
\end{table}

\begin{figure}[htbp]
%\center
\vskip-0.5cm
\vbox{\quad\psfig{figure=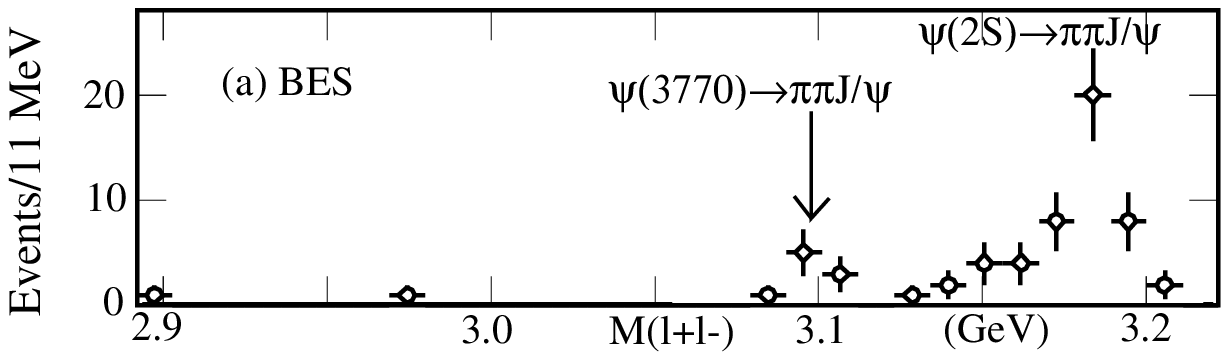,width=7.0truecm,height=4.0truecm}}
\vskip-0.8cm
\vbox{\psfig{figure=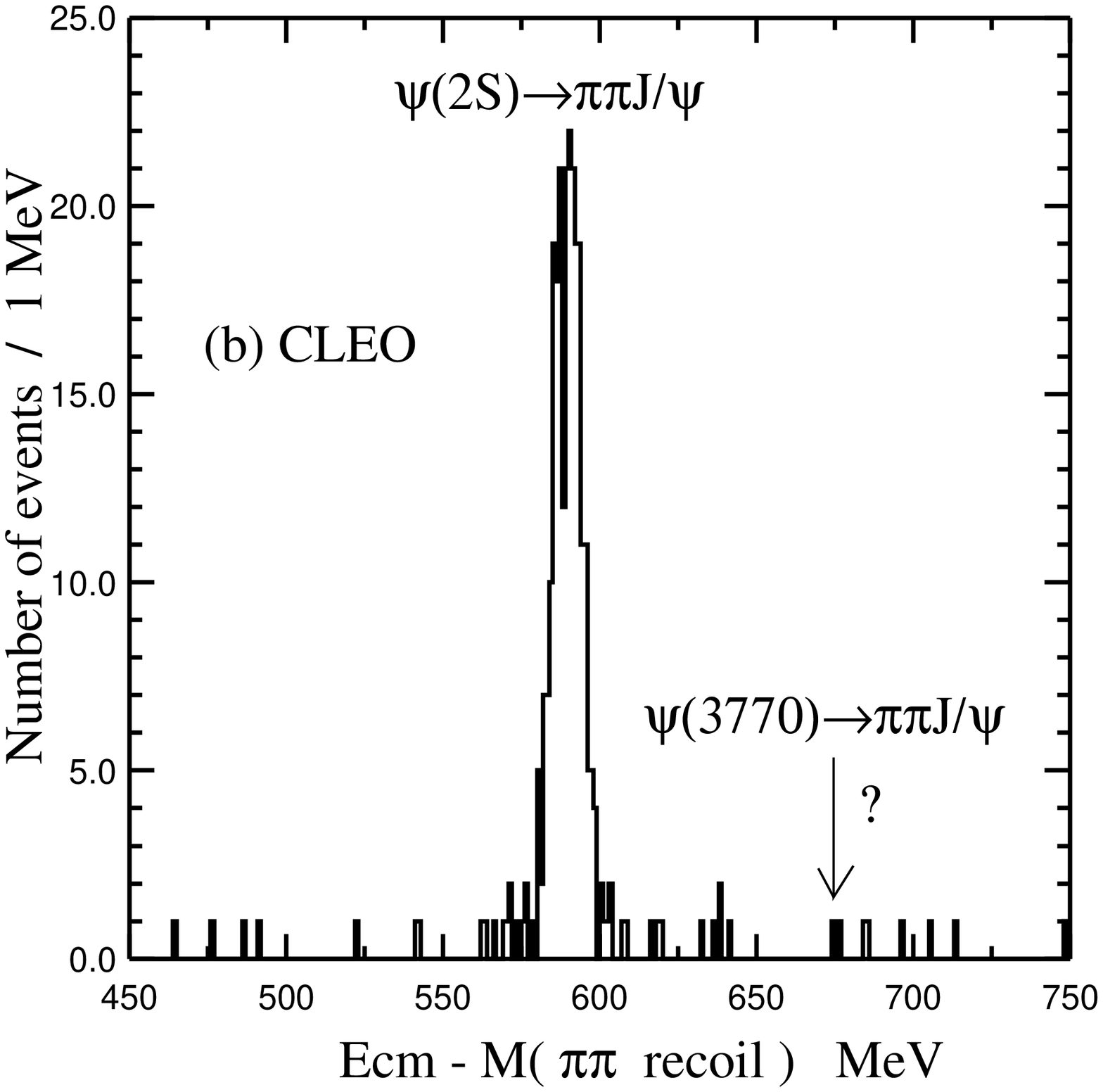,width=8.0truecm,height=4.5truecm}}
\vskip-0.3cm
\caption{Searches for $\psi(3770)\to\pi^+\pi^- J/\psi$ transitions.
         The expected signal positions are indicated with arrows.
         The dominant peak in each distribution is due to
         $\ee\to\gamma\psi(2S)$, $\psi(2S)\to\pi^+\pi^- J/\psi$.
         (a) BES results. $M(l^+l^-)$ mass for kinematically constrained 
         $\pi^+\pi^-l^+l^-$ events is plotted. The signal is expected
         at the $J/\psi$ mass. 
         (b) CLEO results. The difference between the center-of-mass
         energy and the recoil mass against the $\pi^+\pi^-$ pair
         is plotted with no kinematic constraints, but after the cut
         on $M(l^+l^-)$ around the $J/\psi$ mass. 
         The signal is expected at the mass difference between the
         $\psi(3770)$ and the $J/\psi$. 
         }
\label{fig:psipp-pp}
\vskip-0.6cm
%\vspace{-0.5cm}
\end{figure}

There are also new results on $1^3D_1\to\pi^+\pi^-1^3S_1$ 
transitions in the charmonium system.
Here, theoretical situation is complicated by 
mixing of the $\psi(1^3D_1)$ state with the $\psi(2^3S_1)$.
The observed state, $\psi(3770)$, is also above the open
flavor threshold, therefore it has a large width for
decay to $D\bar D$ which suppresses branching ratios for
any transitions to the other charmonium states.
BES claims observation of such a transition in
the data consisting of $5.7\times 10^4$ $\psi(3770)$ decays\rlap.\,\cite{BESpipi}
Their reconstruction efficiency is 17\%. They  
observe 9 signal candidates with an estimated 
background of $2.2\pm0.4$ events (see Fig.~\ref{fig:psipp-pp}a).
The corresponding
branching ratio is $(0.59\pm0.26\pm0.16)\%$.
Such a large branching ratio would favor 
the Kuang-Yan model\cite{KuangYanPsipp}
contrary to the $\bb$ results discussed above.
CLEO-c has also analyzed their first $\psi(3770)$
sample\rlap.\,\cite{CLEOinternal} 
They have a smaller sample of $\psi(3770)$
decays ($4.5\times 10^4$) but larger reconstruction
efficiency (37\%). They have only 2 events in the
signal region, consistent with the estimated background
level (see Fig.~\ref{fig:psipp-pp}b).
They set a preliminary upper limit 
$\B(\psi(3770)\to\pi^+\pi^- J/\psi)<0.26\%$ (90\% CL).
This result is not inconsistent with the BES value because
of the large experimental errors in both measurements.
However, the CLEO result indicates that 
it is premature to favor the Kuang-Yan model on the basis
of the $\psi(3770)$ data. BES is already analyzing a larger
data sample and CLEO is expected to increase their statistics 
by an order of magnitude this fall, thus more accurate results
are expected next year.

%\section{Annihilation Decays}

\vskip-0.2cm
\section{New Particle Discovered by BELLE}

\begin{figure}[htbp]
\center
\vskip-0.1cm
\psfig{figure=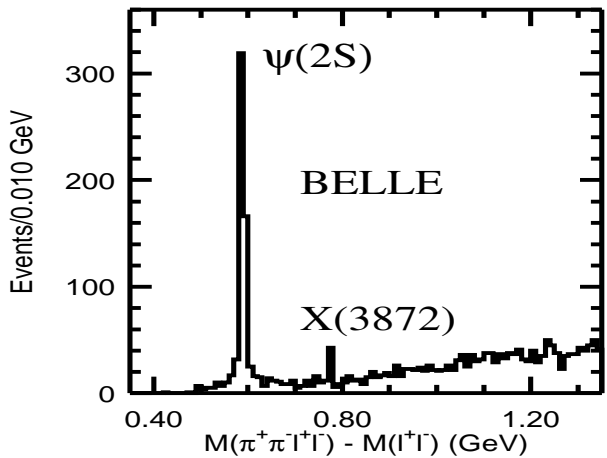,width=8.0truecm,height=6.0truecm}
\vskip-0.1cm
\caption[]{
   Distribution of $M(\pi^+\pi^-l^+l^-)-M(l^+l^-)$ 
   as observed by BELLE for $B$ decay candidates\rlap.\,\cite{BELLEX}
   The first peak corresponds to the $\psi(2S)$ resonances.
   The second peak represents the $X(3872)$ particle.}
\label{fig:BELLESignal}
\vskip-0.6cm
\end{figure}

Decays to $\pi^+\pi^- J/\psi$ are also a way to search
for other charmonium states, for example $h_c(1^1P_1)$,
$\psi(1^3D_2)$, $\psi(1^1D_2)$ etc..
Many of these states cannot be directly formed
in $\ee$ annihilation, however they can be produced
in $\bar pp$ annihilation, $B$ decays or
fragmentation processes.
BELLE inspected resonance structures in
the $\pi^+\pi^- J/\psi$ system produced in the decay
$B^\pm\to K^\pm (\pi^+\pi^- J/\psi)$,
$J/\psi\to l^+l^-$.
The distribution of $\Delta M=M(\pi^+\pi^- J/\psi)-M(J/\psi)$ observed
by BELLE\cite{BELLEX}
in a sample of  $3\times 10^8$ $B$ mesons 
is shown in Fig.~\ref{fig:BELLESignal}.
The prominent peak is due to the $\psi(2^3S_1)$.
There is also a smaller but very significant peak
observed at a higher mass.
By performing an unbinned maximum likelihood fit to
$\Delta M$, the beam-constraint $B$ meson mass
and energy difference between the $B$ candidate and
the beam energy 
BELLE finds $35.7\pm6.8$ events in the second peak with
a statistical significance of 10.3 standard deviations.
The mass determined from the $\Delta M$ peak position is
$(3872.0\pm0.6\pm0.5)$ MeV. 
The observed width of the peak is consistent with the
detector resolution, therefore the new state is long-lived
($\Gamma_{tot}<2.3$ MeV at 90\%\ CL).

\begin{figure}[htbp]
\center
\vskip-1.2cm
\psfig{figure=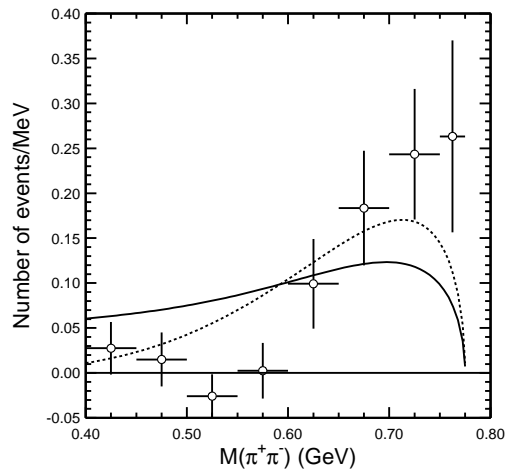,width=7.0truecm}
\vskip-0.1cm
\caption[]{
   Dipion mass distribution in BELLE's data\cite{my-mpp} 
   for $X(3872)\to\pi^+\pi^- J/\psi$.
   The data are compared to the theoretical shapes
   predicted by Yan\cite{Yan80} for $D\to S$ (solid line)
   and $S\to S$ (dashed line) transitions in the quarkonium
   system. The distributions were normalized to the same
   number of events.
   The efficiency dependence on $M(\pi^+\pi^-)$ has been 
   neglected in the comparison.}
\label{fig:BELLE-mpp}
\vskip-0.6cm
%\vspace{-0.5cm}
\end{figure}

The
invariant mass of the $\pi^+\pi^-$ system for
the signal events is strongly peaked at
high mass values.
The peaking is stronger than 
predicted by Yan\cite{Yan80}
with the multipole expansion 
approach for $S\to S$ transitions, 
and much stronger than predicted for $D\to S$
transitions in the quarkonium system, 
as illustrated in 
Fig.~\ref{fig:BELLE-mpp}.
The observed dipion mass distribution
is suggestive of the isospin violating
$X(3872)\to\rho^0  J/\psi$ process
instead.
Isospin violation can be 
experimentally verified
by measuring 
$\Gamma(X(3872)\to\pi^0\pi^0  J/\psi)/
 \Gamma(X(3872)\to\pi^+\pi^-  J/\psi)$\rlap.\,\cite{VoloshinX,BarnesGodfrey}
For isospin conserving $\pi\pi$ transitions, this
ratio should be approximately 1/2, whereas $\rho^0$ does
not decay to $\pi^0\pi^0$.

Soon after BELLE's announcement 
at the Lepton-Photon Symposium, the $X(3872)$ particle
was confirmed by CDF in 
$\pi^+\pi^- J/\psi$, $J/\psi\to\mu^+\mu^-$\rlap.\,\cite{cdfX}
In 220 pb$^{-1}$ of Run-II data 
they observe $704\pm67$ signal events.
The preliminary mass measurement,
$3871.4\pm0.7\pm0.4$ MeV,
is consistent with BELLE's result. 
Their data also show peaking at high
dipion mass.

Perhaps the most puzzling property of the newly
discovered particle is its mass.
Within the experimental errors it coincides with 
the sum of $D^0$ and $D^{*0}$ masses: 
$m_X - m_{D^0} - m_{D^{*0}}= 0.2\pm0.9$ MeV (using the 
average of the BELLE and CDF results).
One exciting interpretation, which actually predicts
this to be the case, 
is a $D\bar D^{*}$ ($+\bar DD^{*}$) molecule. 
Constituent mesons would likely be in a relative S-wave
creating a $J^P=1^+$ state.
``Molecular charmonium'' was first
discussed in the literature in the mid-seventies\cite{molcharm} 
and was initially introduced to explain the
complicated coupling of the $\ee$ resonances
above the open flavor threshold to various
decay modes involving $D$ and $D^*$ meson pairs.
A satisfactory description of the $\ee$ data
was later achieved within a simple $\cc$ bound state
model\rlap.\,\cite{Eichten80,Bradley}
However, since meson molecules were also
proposed in the context of 
light hadron spectroscopy\rlap,\,\cite{lightMolecules}
the concept of molecular charmonium did not go
away. Interactions in the $D\bar D^*$ system
(also $B\bar B^*$) were found to be attractive
when described by the pion-exchange 
force\rlap.\,\cite{Tornqvist,otherPionEx}
No such force would exist in the
$D\bar D$ (or $B\bar B$) system.
Quantitative estimates showed that the  $D\bar D^*$ system
could be only loosely bound if at all, with better
prospects for the $B\bar B^*$ molecule.
Tornqvist showed that in the limit of isospin symmetry
only an isoscalar molecule would be bound\rlap.\,\cite{Tornqvist} 
Isospin is expected to be broken since the binding 
energy is comparable to the isospin mass splittings
(the $D^+D^{-*}$ threshold is 8 MeV above the
$D^0\bar D^{0*}$ threshold).
Close and Page\cite{ClosePage}
argued that the isospin breaking
does not change the conclusion that there is only one
molecular system expected near the $D\bar D^*$ threshold.
The loose binding makes it difficult for the molecule
to rearrange the quark content in the meson subcomponents,
thus strong decays to a charmonium state plus light
hadrons are expected to have small widths.
Decays to $\pi^+\pi^- J/\psi$ could proceed via
the isospin violating $\rho^0 J/\psi$ channel
consistent with the BELLE's dipion mass
distribution. 
The molecular interpretation predicts that
decays to $D^0 \bar D^0 \pi^0$ and 
$D^0 \bar D^0 \gamma$ should occur,
since the $D^{0*}$ is almost on shell.
The ratio of the widths for these decays should
be approximately 3:2 and their sum 
60-100 keV\rlap.\,\cite{Tornqvist,ClosePage,BraatenKusunoki,BarnesGodfrey} 
Therefore, the molecular model 
is consistent with the narrow width of
the observed state.

Traditional charmonium states can also be
narrow at 3872 MeV if they cannot decay to
$D\bar D$. Possible candidates, ordered according
to increasing mass predictions are
$\psi(1^3D_{2^{--}})$, $\eta_{c2}(1^1D_{2^{-+}})$,
$\chi_c(2^3P_{1^{++}})$, $h_c(2^1P_{1^{+-}})$ and
$\eta_c(3^1S_{0^{-+}})$.

The lower excitations of the $\eta_c$ are 
already broader than the upper limit on the
natural width of the new state, thus we can
discard the $\eta_c(3^1S_{0^{-+}})$ possibility. 
The other listed
states can satisfy this upper limit.
For the remaining options,
radiative E1 transitions to the other
charmonium states should dominate 
over $\pi\pi J/\psi$. This is especially
true for the singlet states, for which
the dipion transition would involve
chromomagnetic interactions (spin-flip). 
Since the $h_c(1^1P_{1^{+-}})$ has not been seen
in $B$ decays, the $h_c(2^1P_{1^{+-}})$
hypothesis is unlikely. 

BELLE has also just observed $B^\pm\to K^\pm\psi(3770)$
($\psi(3770)\to D\bar D$) for the first time\rlap.\,\cite{BelleKpsipp}
The measured branching ratio, $(4.8\pm1.1\pm1.2)\times 10^{-4}$,
is comparable to\cite{PDG2002}
$\B(B^\pm\to K^\pm\psi(2S))=(6.8\pm0.4)\times 10^{-4}$.
Theoretically, $\B(B^\pm\to K^\pm\psi(1^3D_2))$ 
is expected to be 1.6 times
larger than the production of $\psi(3770)$ 
(assumed to be $\psi(1^3D_1)$)\rlap.\,\cite{YuanQiaoChao97}
Together, with BELLE's result\rlap,\,\cite{BELLEX}
$\B(B^\pm\to K^\pm X)\B(X\to\pi^+\pi^-J/\psi)$
$=(0.063\pm0.012\pm0.007)$
$\times\B(B^\pm\to K^\pm \psi(2S))\B(\psi(2S)\to\pi^+\pi^-J/\psi)$,
the $\psi(1^3D_2)$ interpretation requires
$\B(\psi(1^3D_2)\to\pi^+\pi^- J/\psi)$ 
to be about 1 to 3\%, much smaller than 
the na\"{\i}ve expectations\cite{EichtenLaneQuigg,BarnesGodfrey} 
for the dominant radiative decay of this state, 
$\B(\psi(1^3D_2)\to\gamma\chi_c(1^3P_1))>50\%$.
Also, as discussed in the previous section 
the $\Upsilon(1^3D_{2})$ state was observed
via its radiative decay, whereas only an upper limit
on the decay to $\pi^+\pi^- \Upsilon$ exists.
However, BELLE observes no evidence for the photon
transitions to $\chi_c(1^3P_1)$ and sets the following 
90\%\ CL\ limit:
$\B(X(3872)\to\gamma\chi_c(1^3P_1))/$
$\B(X(3872)\to\pi^+\pi^- J/\psi(1^3S_1))$
$<0.89$.
Coupled channel effects can be big
(proximity of the $D\bar D^*$ threshold!) 
and significantly alter the 
na\"{\i}ve predictions for the radiative widths\rlap.\,\cite{EichtenQWG} 
Quantitative estimates are needed.

\begin{figure}[htbp]
\center
\vskip-1.2cm
\psfig{figure=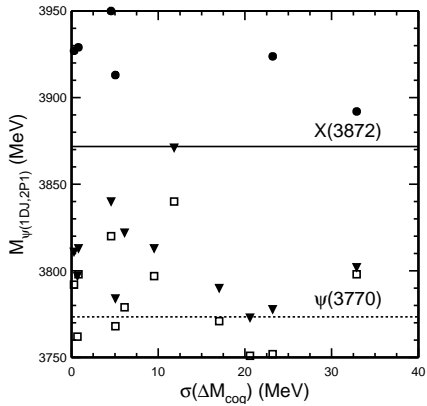,width=6.0truecm}
\caption[]{The predicted masses (vertical axis) of the
 $\psi(1^3D_1)$ (squares), $\psi(1^3D_2)$ (triangles) and
 $\chi_c(2^3P_1)$ (circles) states 
 versus overall quality
 of the potential model
 expressed as RMS of the differences for
 the center-of-gravity masses of long-lived 
 $\cc$ states (horizontal axis).
 The measured masses of $\psi(3770)$ and $X(3872)$
 are indicated by horizontal bars.
 Predictions for the 1S,2S and 1P masses are included
 in the RMS calculation 
 for all models displayed here\rlap.\,\cite{p1d2pot}
% See also 
% a similar plot for the $\bb$ system is shown in Fig.~\protect\ref{fig:bbdmass}.
 }
\label{fig:ccdmass}
\vskip-0.6cm
%\vspace{-0.5cm}
\end{figure}

It will also be
useful to resolve the experimental controversy
about $\B(\psi(3770)\to\pi^+\pi^- J/\psi)$, since
the width for
$\psi(1^3D_2)\to\pi^+\pi^- J/\psi$ 
is related to the width for
$\psi(1^3D_1)\to\pi^+\pi^- J/\psi$.
The CLEO-c upper limit 
on the $\psi(3770)\to\pi^+\pi^- J/\psi$ branching
ratio (see the previous section)
suggests that the latter is
not necessarily much larger than the value
induced by the $2^3S_1-1^3D_1$ mixing,
while the BES measurement indicates a rather
large rate for direct $\pi\pi$ transitions
between the charmonium triplet 
$1D$ and $1S$ states. 
The BES result implies\cite{BarnesGodfrey} 
$\B(\psi(1^3D_2)\to\pi^+\pi^- J/\psi)\sim(20-40)\%$
which would make it easier to accommodate
BELLE's results in the $1^3D_2$ interpretation of 
the $X(3872)$.
The observed dipion mass distribution does
not fit the shape for $1D_2\to 1S$
transitions predicted by the multipole
expansion model (see Fig.~\ref{fig:BELLE-mpp}).
However, some dynamical effects beyond
this model can alter the dipion distribution.
Finally, the $\psi(1^3D_2)$ interpretation 
is disfavored by the mass predictions from 
potential model. This is illustrated 
in Fig.~\ref{fig:ccdmass}.
All potential models but one predict the 
$\psi(1^3D_2)$ mass to be about 70 MeV
lower then the $X(3872)$ mass.
The model by Fulcher\cite{Fulcher91}
predicts this mass right at the observed value,
however it overestimates the $\psi(1^3D_1)$
mass by similar amount. In other words, 
none of the existing calculations can accommodate
the  $X(3872)$ and $\psi(3770)$ as spin 2 and 1
members of the $1^3D_J$ triplet. 
Coupled channel effects and $1^3D_1-2^3S_1$
mixing
can increased the mass splitting relatively to the
na\"{\i}ve potential model calculations.
Quantitative estimates of these effects are needed.

The mass of the $\chi_c(2^3P_{1^{++}})$ state
is predicted by the potential models to be
higher than the $X(3872)$ mass
(see Fig.~\ref{fig:ccdmass}).
Thus, significant coupled channel effects
would need to be at work for this interpretation
as well.
Decays of $B^\pm$ to $K^\pm\chi_c(1^3P_{1})$
are observed with a rate 
comparable to 
$K^\pm\psi(2S)$ and $K^\pm J/\psi(1S)$\rlap.\,\cite{PDG2002}
Therefore, decays to $K^\pm\chi_c(2^3P_{1})$ 
should also occur. 
If the $X(3872)$ is the $\chi_c(2^3P_{1})$ state
then the ratio
$\B(X(3872)\to\gamma J/\psi(1^3S_1))/$
$\B(X(3872)\to\pi^+\pi^- J/\psi(1^3S_1))$
should be large. 
The photon transition here
is of an E1 type and is observed with a 21\%\
branching ratio in the $\bb$ system. 
Barnes and Godfrey predict 3.5\%\ 
for the $\chi_c(2^3P_{1})$ state\rlap.\,\cite{BarnesGodfrey}
The $\chi_c(2^3P_{1})\to\pi^+\pi^- J/\psi(1^3S_1)$ 
transition is of an isospin violating type
and the branching ratio must be small even in the
charmonium system.
The BELLE experiment does not see
evidence for significant
$X(3872)\to\gamma J/\psi(1^3S_1)$
branching ratio. Quantitative results should
be soon available\rlap.\,\cite{BELLEPrivate}

Finally, $X(3872)$ could be a $c\bar c g$
hybrid state\rlap,\,\cite{ClosePage,BarnesGodfrey} though
the masses of the hybrid states are predicted
to be significantly 
higher than the observed mass.

The molecular, conventional charmonium or
hybrid charmonium interpretations of
the $X(3872)$ should not be viewed as
mutually exclusive options.
For example, the production rate for
molecular charmonium in $B$ decays
is expected to be small.
However, the production rate can
be enhanced by mixing of the molecular
system with conventional charmonium
states\cite{PakvasaSuzuki,ClosePage,BraatenKusunoki}
(e.g. with $\chi_c(2^3P_1)$ for the $1^{++}$ molecule).
Mixing would influence the pattern of decay branching
ratios as well\rlap.\,\cite{PakvasaSuzuki,ClosePage,BarnesGodfrey}
%... relevance of D_sJ, BES gamma p pbar ?

More experimental studies of $X(3872)$
production and decays are needed
to clarify nature of this state.

\vskip-0.2cm
\section{Summary and Outlook}

Heavy quarkonium physics has been recently experimentally
revitalized. Large data samples collected
in $e^+e^-$ and $\bar pp$ annihilation 
by BES-II ($\cc$), CLEO-III ($\bb$)
and E835 ($\cc$) are still being analyzed.
The $B$-meson gateway to charmonium is now wide
open with $3\times 10^8$ $B$ decays recorded
by BELLE and BABAR.
There has been similar progress in
theory by the development of NRQCD and
much improved calculations with 
QCD on the lattice.

More results for charmonium are expected 
in the future. In addition to more results
from the BES-II experiment, the CLEO-c program
will contribute greatly. The first CLEO-c
results have been presented.
Run-II data from CDF are proving
to be useful as well. In the farther future,
BTeV and LHCb are likely to contribute
to quarkonium physics.
The BES-III/BEPC-II project was recently
approved in China.
There is also a proposal for a new dedicated
$\bar pp$ machine to explore charmonium
physics (PANDA at GSI).

Prospects for more $\bb$ data are not
as well defined. In principle, CLEO
at CESR could go back to the high energy
running and accumulate more statistics
for the narrow $\Upsilon$ resonances.
BELLE and BABAR could accumulate 
$\Upsilon$ data with even higher rates
if their $B$-factories are ever 
utilized to produce these states.

The discovery of the $X(3872)$ particle 
by BELLE, and its likely compound nature,
underlines the importance of
the heavy quarkonium systems. Previously
charmonium played a crucial role in the
solidification of the na\"{\i}ve quark model.
It will not be a big surprise if
it provides the first convincing proof for
the existence of hadronic systems 
beyond the na\"{\i}ve quark model.

\vskip-0.2cm
\section{Acknowledgments}

I would like to thank many colleagues from
BABAR, BELLE, BES, CDF, CLEO, E835, H1 and
ZEUS for providing me with information
contained in this report.
I apologize for not being able to discuss 
all the recent experimental results because
of the length limitations.
I would also like to thank 
Steve Godfrey, Jonathan Rosner, Mikhail Voloshin and
Tung-Mow Yan for discussions concerning
theoretical aspects.
Special thanks to Sheldon Stone for
help in refining this manuscript.


\vskip-0.2cm
\begin{thebibliography}{99}

\bibitem{GrossmanLP03}
Y.~Grossman,
``Beyond Standard Model
sensitivity in K and B Physics'',
talk presented at Lepton Photon 2003.
%Fermilab, Batavia, IL, USA,
%11-16 August 2003.

\bibitem{LepageLP03}  
P.~Lepage, 
``Lattice Gauge Theory'',
talk presented at Lepton Photon 2003.
%Fermilab, Batavia, IL, USA,
%11-16 August 2003.

\bibitem{NRQCD}  
W.~E.~Caswell, G.~P.~Lepage, {\em Phys.\ Lett.}\ {\bf B167}, 437 (1986);
B.~A.~Thacker, G.~P.~Lepage, {\em Phys.\ Rev.}\ {\bf D43}, 196 (1991);
G.~T.~Bodwin, E.~Braaten, G.~P.~Lepage, {\em Phys.\ Rev.}\ {\bf D51}, 1125 (1995);
{\it erratum} {\bf D55}, 5853 (1997);
N.~Brambilla, A.~Pineda, J.~Soto, A.~Vairo,
{\em Nucl.\ Phys.}\ {\bf B566}, 275 (2000).


\bibitem{NRQCDrev}
For a recent review see:
M.~Kr\"amer, {\em Prog.\ Part.\ Nucl.\ Phys.}\ {\bf 47}, 141 (2001).

\bibitem{CS}
E.~L.~Berger, D.~Jones, {\em Phys.\ Rev.}\ {\bf D23}, 1521 (1981);
R.~Baier, R.~R\"uckl, {\em Phys.\ Lett.}\ {\bf B102}, 364 (1981).

\bibitem{CEM}
H.~Fritzsch, {\em Phys.\ Lett.} {\bf B67}, 217 (1977);
F.~Halzen, {\em Phys.\ Lett.} {\bf B69}, 105 (1977);
M.~Gl\"uck, J.~F.~Owens, E.~Reya, {\em Phys. Rev.} {\bf D17}, 2324 (1978);
G.~A.~Schuler, R.~Vogt, {\em Phys.\ Lett.} {\bf B387}, 181 (1996);
J.~F.~Amundson, O.~J.~Eboli, E.~M.~Gregores, F.~Halzen,
   {\em Phys.\ Lett.} {\bf B390}, 323 (1997).

\bibitem{H1ZEUS}
C.~Adloff \etal\ (H1),
%Papers contributed to Lepton Photon 2003 (Abstracts 151,156,157,162,188);
papers contributed to Lepton-Photon 2003: ``Diffractive
                                           Photoproduction of
                                           $\psi$(2S) Mesons at
                                           HERA'', $\sharp$151;
``Inelastic
                                    Leptoproduction of
                                    J/$\psi$ Mesons at
                                    HERA'', $\sharp$156;
``Inelastic
                                           Photoproduction of J/$\psi$
                                           Mesons at HERA'', $\sharp$157;
``Diffractive
                                           Photoproduction of J/$\psi$
                                           Mesons with Large
                                           Momentum Transfer at
                                           HERA'', $\sharp$162;
``Elastic
                                      Photoproduction of
                                      J/$\psi$ Mesons'', $\sharp$188;
S.~Chekanov \etal\ (ZEUS),
%Papers contributed to Lepton Photon 2003 (Abstracts $\sharp$91,$\sharp$106,$\sharp$109,$\sharp$111).
papers contributed to Lepton-Photon 2003: ``Measurements of
                                           inelastic J/$\psi$ and $\psi'$
                                           photoproduction at
                                           HERA'', $\sharp$91;
``Measurement of
                                        proton-dissociative
                                        diffractive photoproduction
                                        of J/$\psi$ mesons at HERA'', $\sharp$106;
``Measurement of inelastic
                                        J/$\psi$ production in deep
                                        inelastic scattering at
                                        HERA'', $\sharp$109;
``Measurement of inelastic
                                        J/$\psi$ helicity distribution at
                                        HERA'', $\sharp$111.

\bibitem{R704}
C.~Baglin \etal\ (R704), {\em Phys.\ Lett.}\ {\bf B172}, 455 (1986).

\bibitem{E760}
T.~A.~Armstrong \etal\ (E760), {\em Nucl.\ Phys.}\ {\bf B373}, 35 (1992).

\bibitem{E835chimw}
S.~Bagnasco \etal\ (E835), {\em Phys.\ Lett.}\ {\bf B533}, 237 (2002);
N.~Pastrone for the E835
at XXXVIIIth Recontres de Moriond, Les Arcs, Savoie, France, 
March 22-29, 2003, hep-ex/0306032. 

\bibitem{E835rad}
M.~Ambrogiani \etal\ (E835), {\em Phys.\ Rev.}\ {\bf D65}, 052002 (2002);
C.~Patrignani at International Workshop on Heavy Quarkonium
at Fermilab, Batavia, IL, USA, September 20-22, 2003.

\bibitem{E835pzpz}
M.~Andreotti \etal\ (E835), {\em Phys.\ Rev.\ Lett.}\ {\bf 91}, 091801 (2003). 

\bibitem{E835etac}
M.~Ambrogiani \etal\ (E835), {\em Phys.\ Lett.}\ {\bf B566}, 45 (2003). 

\bibitem{E835etacp}
M.~Ambrogiani \etal\ (E835), {\em Phys.\ Rev.}\ {\bf D64}, 052003 (2001).

\bibitem{CBetacp}
C.~Edwards \etal\ (Crystal Ball), {\em Phys.\ Rev.\ Lett.}\ {\bf 48}, 70 (1982).

\bibitem{R704hc}
C.~Baglin \etal\ (R704), {\em Phys.\ Lett.}\ {\bf B171}, 135 (1986).

\bibitem{E760hc}
T.~A.~Armstrong \etal\ (E760), {\em Phys.\ Rev.\ Lett.}\ {\bf 69}, 2337 (1992).

\bibitem{hcRumors}
See e.g. A.~Martin, J.-M.~Richard CERN Cour.\ {\bf 43N3}, 17 (2003).

\bibitem{CesterPrivate} 
Rosanna Cester, private communication.

\bibitem{PDG2002}
K.~Hagiwara \etal\ (PDG), {\em Phys.\ Rev.}\ {\bf D66}, 010001 (2002).

\bibitem{BESetac}
J.~Z.~Bai \etal\ (BES), {\em Phys.\ Lett.}\ {\bf B555}, 174 (2003).
 
\bibitem{BABARetac}
C.~Wagner for BABAR 
at XXXVIIIth Recontres de Moriond, Les Arcs, Savoie, France, 
March 22-29, 2003, hep-ex/0305083;
E.~Robutti for BABAR
at International Europhysics Conference on High Energy Physics
EPS, Aachen, Germany, July 17-23, 2003.

\bibitem{BELLEetac}
F.~Fang \etal\ (BELLE),
{\em Phys.\ Rev.\ Lett.}\ {\bf 90}, 071801 (2003), hep-ex/0208047.

\bibitem{CLEOcetac}
The preliminary CLEO-c result for the $\eta_c$ mass,
$(2976.1\pm2.4\pm3.3)$ MeV, 
comes from the photon energy measurement in 
$\psi'\to\gamma\eta_c$ (see the text). 

\bibitem{BELLEetacp}
S.~K.~Choi \etal\ (BELLE),
{\em Phys.\ Rev.\ Lett.}\ {\bf 89}, 102001 (2002),
Erratum-ibid.\ {\bf 89}, 129901 (2002).

\bibitem{BELLEdoubleccOld} 
K.~Abe \etal\ (BELLE), {\em Phys.\ Rev.\ Lett.}\ {\bf 89}, 142001 (2002);
K.~Abe \etal, BELLE-PREPRINT-2003-7, KEK-PREPRINT-2003-24, hep-ex/0306015.
 
\bibitem{BELLEdoubleccNew} 
K.~Abe \etal\ (BELLE),
%Paper contributed to Lepton Photon 2003 (Abstract 274),
paper contributed to Lepton-Photon 2003, ``A study of double $c\bar{c}$
                                         production in $e^+ e^-$
                                         annihilation at $\sqrt{s}$
                                         $\approx$ 10.6 GeV'', $\sharp$274;

T.~Uglov and R.~Seuster for BELLE collaboration, talks presented
at International Europhysics Conference on High Energy Physics
EPS, Aachen, Germany, July 17-23, 2003.

\bibitem{doubleccPuzzle} 
%
E.~Braaten, J.~Lee, {\em Phys.\ Rev.}\ {\bf D67}, 054007 (2003); 
%
K.-Y.~Liu, Z.-G.~He, K.-T.~Chao, {\em Phys.\ Lett.}\ {\bf B557}, 45 (2003); 
%
G.~T.~Bodwin, J.~Lee, E.~Braaten, {\em Phys.\ Rev.}\ {\bf D67}, 054023 (2003); 
{\em Phys.\ Rev.\ Lett.}\ {\bf 90}, 162001 (2003); 
%
A.~B.~Kaidalov, {\em JETP Lett.}\ {\bf 77}, 349 (2003), 
{\em Pisma Zh.\ Eksp.\ Teor.\ Fiz.}\ {\bf 77}, 417 (2003), hep-ph/0301246; 
%
A.~V.~Berezhnoy, A.~K.~Likhoded, hep-ph/0303145; 
%
K.-Y.~Liu, Z.-G.~He, K.-T.~Chao, {\em Phys.\ Rev.}\ {\bf D68}, 031501 (2003); 
%
K.~Hagiwara, E.~Kou, C.-F.~Qiao, {\em Phys.\ Lett.}\ {\bf B570}, 39 (2003); 
%
A.~V.~Luchinsky, hep-ph/0305253; 
%
S.~J.~Brodsky, A.~S.~Goldhaber, J.~Lee, {\em Phys.\ Rev.\ Lett.}\ {\bf 91}, 112001 (2003); 
%
B.~L.~Ioffe, D.~E.~Kharzeev, hep-ph/0306062; 
%
S.~Fleming, A.~K.~Leibovich, T.~Mehen 
CMU-HEP-03-06, FERMILAB-PUB-03-069-T, hep-ph/0306139;
%
C.-F.~Qiao, J.-X.~Wang, hep-ph/0308244.

\bibitem{CLEOetacp}
J.~Ernst \etal\ (CLEO), CLEO-CONF-03-05, hep-ex/0306060,
paper contributed to Lepton-Photon 2003, ``Observation of ${\eta_c}^{\prime}$
                                          Production in
                                          Two-photon Fusion at
                                          CLEO'', $\sharp$37.


\bibitem{hyperfineRatioPlot}
References are ordered according to the
increasing value of $\RHF$ given in 
square brackets.
\def\1#1{$\left[#1\right]$}
H.~Ito, Prog. of {\em Theor. Phys.}\ {\bf 84}, 94  (1990) \1{0.29};
T.~A.~Lahde, {\em Nucl.\ Phys.}\ {\bf A714}, 183 (2003) \1{0.37};
S.~Godfrey, N.Isgur, {\em Phys.\ Rev.}\ {\bf D32}, 189 (1985) \1{0.46};
S.~N.~Jena, {\em Phys.\ Lett.}\ {\bf B123}, 445 (1983)  \1{0.48};
J.~Carlson, J.~B.~Kogut, V.R.Pandharipande, 
          {\em Phys.\ Rev.}\ {\bf D28}, 2807 (1983) \1{0.55};
T.~A.~Lahde, C.~J.~Nyfalt, D.~O.~Riska 
          {\em Nucl.\ Phys.}\ {\bf A645}, 587 (1999) \1{0.56};
S.~N.~Gupta, W.~W.~Repko, C.~J.~Suchyta III,
          {\em Phys.\ Rev.}\ {\bf D39},  974 (1989) \1{0.61};
D.~Beavis, S.-Y.~Chu, B.~R.~Desai, P.~Kaus, 
          {\em Phys.\ Rev.}\ {\bf D20}, 2345 (1979) \1{0.62};
N.~Barik, S.~N.~Jena, {\em Phys.\ Rev.}\ {\bf D24}, 680 (1981) \1{0.70};
S.~N.~Gupta, S.~F.~Radford, W.~W.~Repko,
   {\em Phys.\ Rev.}\ {\bf D26}, 3305 (1982), 
   {\em Phys.\ Rev.}\ {\bf D30}, 2424 (1984) 
   \1{0.73};
L.~P.~Fulcher, {\em Phys.\ Rev.}\ {\bf D44}, 2079 (1991) \1{0.74};
J.~S.~Kang, {\em Phys.\ Rev.}\ {\bf D20}, 2978 (1979) \1{0.76};
J.~Baacke, Y.~Igarashi, G.~Kasperidus
   {\em Z.\ Phys.}\ {\bf C13}, 131 (1982) \1{0.80};
S.~N.~Gupta, S.~F.~Radford, W.~W.~Repko, 
   {\em Phys.\ Rev.}\ {\bf D34}, 201  (1986) \1{0.83};
D.~Ebert, R.~N.~Faustov, and V.~O.~Galkin
   {\em Phys.\ Rev.}\ {\bf D67}, 014027 (2003) \1{0.84}.

\bibitem{BadalianBakker}
A.~M.~Badalian, B.~L.~G.~Bakker, {\em Phys.\ Rev.}\ {\bf D67}, 071901 (2003).%0.48+-0.08
 
\bibitem{RecksiegelSumino}
S.~Recksiegel, Y.~Sumino,
TU-691, TUM-HEP-508-03,  hep-ph/0305178.%0.42

\bibitem{LatticeHF}
M.~Okamoto \etal\ (CP-PACS Collaboration)
{\em Phys.\ Rev.}\ {\bf D65}, 094508 (2002).%0.50
  
\bibitem{LEPetab}
A.~Heister \etal\ (ALEPH), {\em Phys.\ Lett.}\ {\bf B530}, 56 (2002); 
M.~Chapkine \etal\ (DELPHI), 
paper contributed to Lepton-Photon 2003, ``Search for $\eta_b$ in
                                           two-photon collisions
                                           with the DELPHI
                                           detector'', $\sharp$62.

%%% to balance text in the two columns
%\balance

\bibitem{m1preds}
The following sample of potential model predictions for the M1 matrix
elements (ordered according to the publication date)
is displayed in Fig.~\protect\ref{fig:m1}:
%  ZB83 GOS84A GOS84B ZSG91A ZSG91B EFG03 Lah03A Lah03B
V.~Zambetakis, N.Byers, {\em Phys.\ Rev.}\ {\bf D28}, 2908 (1983); %ZB83
H.~Grotch, D.~A.~Owen, K.~J.~Sebastian, {\em Phys.\ Rev.}, D30, 1924 (1984)
(2 entries: scalar and vector confinement potential); %GOS84A  %GOS84B
%GI85A,B
S.~Godfrey, N.~Isgur, {\em Phys.\ Rev.}\ {\bf D32}, 189 (1985)
(2 entries: based on quoted transition moments and wave functions
 respectively);
X.~Zhang, K.~J.~Sebastian, H.~Grotch, {\em Phys.\ Rev.}\ {\bf D44}, 1606 (1991)
(2 entries: scalar-vector and pure scalar confinement potential); %ZSG91A,B
D.~Ebert, R.~N.~Faustov, V.~O.~Galkin {\em Phys.\ Rev.}\ {\bf D67}, 014027 (2003); %EFG03
T.~A.~Lahde, {\em Nucl.\ Phys.} A714, 183(2003)
(2 entries: without and with exchange currents).%Lah03A Lah03B
Values of the M1 matrix elements in the $\bb$ system are
displayed for the photon energies and $b$ quark mass assumed in
S.~Godfrey, J.~L.~Rosner, {\em Phys.\ Rev.}\ {\bf D64}, 074011 (2001),
Erratum-ibid.\ {\bf D65}, 039901 (2002).
 
\bibitem{e1preds}
The following sample of potential model predictions for the E1 matrix
elements (ordered according to the publication date)
is displayed in Fig.~\protect\ref{fig:e1}:
D.~Pignon, C.~A.~Piketty, {\em Phys.\ Lett.}\ {\bf 74B}, 108 (1978);   
E.~Eichten, K.~Gottfried, T.~Kinoshita, K.~D.~Lane, T.~M.~Yan,
{\em Phys.\ Rev.}\ {\bf D21}, 203 (1980);
W.~Buchmuller, G.~Grunberg, S.-H.~Tye
{\em Phys.\ Rev.\ Lett.}\ {\bf 45}, 103 (1980), {\em Phys.\ Rev.}\ {\bf D24}, 132 (1981);
C.~Quigg, J.~L.~Rosner, {\em Phys.\ Rev.}\ {\bf D23}, 2625 (1981) 
(2 entries: $\cc$, $\bb$ potential respectively);
J.~Baacke, Y.~Igarashi, G.~Kasperidus, {\em Z. Phys.}\ {\bf C13}, 131 (1982);
R.~McClary, N.~Byers, {\em Phys.\ Rev.}\ {\bf D28}, 1692 (1983);
P.~Moxhay, J.~L.~Rosner, {\em Phys.\ Rev.}\ {\bf D28}, 1132 (1983);
H.~Grotch, D.~A.~Owen, K.~J.~Sebastian, {\em Phys.\ Rev.}\ {\bf D30}, 1924 (1984);
S.~N.~Gupta, S.~F.~Radford, W.~W.~Repko, 
  {\em Phys.\ Rev.}\ {\bf D26}, 3305 (1982), 
  {\em Phys.\ Rev.}\ {\bf D30}, 2424 (1984);
S.~N.~Gupta, S.~F.~Radford, W.~W.~Repko, {\em Phys.\ Rev.}\ {\bf D34}, 201 (1986);
M.~Bander, D.~Silverman, B.~Klima, U.~Maor,
  {\em Phys.\ Lett.}\ {\bf B134}, 258 (1984), {\em Phys.\ Rev.}\ {\bf D29}, 2038 (1984),
  {\em Phys.\ Rev.}\ {\bf D36}, 3401 (1987);
W.~Kwong, J.~L.~Rosner, {\em Phys.\ Rev.}\ {\bf D38}, 279 (1988);
L.~P.~Fulcher, {\em Phys.\ Rev.}\ {\bf D37}, 1259 (1988);
S.~N.~Gupta, W.~W.~Repko, C.~J.~Suchyta III, {\em Phys.\ Rev.}\ {\bf D39},  974 (1989);
L.~P.~Fulcher, {\em Phys.\ Rev.}\ {\bf D42}, 2337 (1990);
A.~K.~Grant, J.~L.~Rosener, E.~Rynes, {\em Phys.\ Rev.}\ {\bf D47}, 1981 (1993);
T.~A.~Lahde, {\em Nucl.\ Phys.}\ {\bf A714}, 183(2003).


\bibitem{Gottfried}
K.~Gottfried, {\em Phys.\ Rev.\ Lett.}\ {\bf 40}, 598 (1978);
and Yan\rlap.\,\cite{Yan80}

\bibitem{Yan80}
T.~M.~Yan, {\em Phys.\ Rev.}\ {\bf D22}, 1652 (1980).

\bibitem{EichtenGottfried77}
E.~Eichten, K.~Gottfried, {\em Phys.\ Lett.}\ {\bf B66}, 286 (1977).

\bibitem{chiomega}
H.~Severini \etal\ (CLEO), CLEO-CONF-03-06,
contributed to 
Lepton Photon 2003,
%Fermilab, Batavia, IL, USA,
%11-16 August 2003,
hep-ex/0307034 

\bibitem{VoloshinOmega}
M.~B.~Voloshin, Mod.\ {\em Phys.\ Lett.}\ {\bf A18}, 1067 (2003). 

\bibitem{YanKuang81}
T.~M.~Yan, Y.~P.~Kuang, {\em Phys.\ Rev.}\ {\bf D24}, 2874 (1981).

\bibitem{NSZV}
M.~B.~Voloshin {\em Nucl.\ Phys.}\ {\bf B154}, 365 (1979);
M.~B.~Voloshin, V.~Zakharov {\em Phys.\ Rev.\ Lett.}\ {\bf 45}, 688 (1980);
V.~A.~Novikov, M.~A.~Shifman, {\em Z.\ Phys.}\ {\bf 8}, 43 (1981);
M.~B.~Voloshin {\em Pisma Zh.\ Eksp.\ Teor.\ Fiz.}
{\bf 37}, 58 (1983) [JETP Lett. 37, 69 (1983)].

\bibitem{Y1DAmsterdam}
S.~E.~Csorna \etal,
CLEO-CONF-02-06,
Contributed to 31st 
International Conference 
on High Energy Physics (ICHEP 2002), 
Amsterdam, The Netherlands, 24-31 Jul 2002,
hep-ex/0207060.

\bibitem{CLEOinternal}
CLEO collaboration, private communication.

\bibitem{y1d2pot}
The following sample of potential models
(ordered according to the increasing
value of $\sigma(\Delta M_{cog})$)
is displayed in Fig.~\protect\ref{fig:bbdmass}:
D.~Ebert, R.~N.~Faustov, V.~O.~Galkin, {\em Phys.\ Rev.}\ {\bf D62}, 034014 (2000);
W.~Kwong, J.~L.~Rosner, {\em Phys.\ Rev.}\ {\bf D38}, 279 (1988);
L.~P.Fulcher, {\em Phys.\ Rev.}\ {\bf D42}, 2337 (1990);
P.~Moxhay, J.~L.~Rosner, {\em Phys.\ Rev.}\ {\bf D28}, 1132 (1983);
L.~P.~Fulcher, {\em Phys.\ Rev.}\ {\bf D44}, 2079 (1991);
S.~Godfrey, N.~Isgur, {\em Phys.\ Rev.}\ {\bf D32}, 189 (1985);
E.~Eichten, F.~Feinberg, {\em Phys.\ Rev.}\ {\bf D23}, 2724 (1981);
T.~A.~Lahde, {\em Nucl.\ Phys.}\ {\bf A714}, 183 (2003);
J.~R.~Hiller, {\em Phys.\ Rev.}\ {\bf D30}, 1520 (1984);
W.~Buchmuller, {\em Phys.\ Lett.}\ {\bf B112}, 479 (1982);
T.~A.~Lahde, C.~J.~Nyfalt, D.~O.~Riska, {\em Nucl.\ Phys.}\ {\bf A645}, 587 (1999);
J.~S.~Kang, {\em Phys.\ Rev.}\ {\bf D20}, 2978 (1979);
M.~Hirano, {\em Prog.\ of Theor.\ Phys.}\ {\bf 83}, 575 (1990);
J.~Baacke, Y.~Igarashi, G.~Kasperidus, {\em Z.\ Phys.}\ {\bf C13}, 131 (1982).

\bibitem{GodfreyRosner}
S.~Godfrey, J.~L.~Rosner
{\em Phys.\ Rev.}\ {\bf D64}, 097501 (2001),
Erratum-ibid.\ {\bf D66}, 059902 (2002). 

\bibitem{RosnerY1Dpipi}
J.~L.~Rosner, {\em Phys.\ Rev.}\ {\bf D67}, 097504 (2003)

\bibitem{Moxhay88}
P.~Moxhay, {\em Phys.\ Rev.}\ {\bf D37}, 2557 (1988).

\bibitem{Ko93}
P.~Ko, {\em Phys.\ Rev.}\ {\bf D47}, 208 (1993).

\bibitem{VoloshinEta}
M.~B.~Voloshin, {\em Phys.\ Lett.}\ {\bf B562}, 68 (2003).

\bibitem{BESpipi}
J.~Z.~Bai {\it et al.} (BES)
hep-ex/0307028.

\bibitem{KuangYanPsipp}
Y.~P.~Kuang, {\em Phys.\ Rev.}\ {\bf D65}, 094024 (2002);
Y.~P.~Kuang, T.M.Yan,
{\em Phys.\ Rev.}\ {\bf D41}, 155 (1990).

\bibitem{BELLEX}
S.~K.~Choi \etal\ (BELLE), hep-ex/0309032;
K.~Abe \etal\ (BELLE), hep-ex/0308029, 
paper contributed to Lepton-Photon 2003, ``Observation of a new
                                           narrow charmonium
                                           state in exclusive
                                           $B^\pm \to K^\pm
                                           \pi^+\pi^- J/\psi$
                                           decays'', $\sharp$307.

\bibitem{VoloshinX}
M.~B.~Voloshin, hep-ph/0309307.

\bibitem{BarnesGodfrey}
T.~Barnes, S.~Godfrey, hep-ph/0311162. 


\bibitem{cdfX}
G.~Bauer (CDF),
at International Workshop on Heavy Quarkonium
at Fermilab, Batavia, IL, USA, September 20-22, 2003. 

\bibitem{molcharm}
M.~Bander, G.~L.~Shaw, P.~Thomas, {\em Phys.\ Rev.\ Lett.}\ {\bf 36}, 695 (1976);
M.~B.~Voloshin, L.~B.~Okun {\em JETP Lett.}\ {\bf 23}, 333 (1976), Pisma
{\em Zh.\ Eksp.\ Teor.\ Fiz.}\ {\bf 23}, 369 (1976);
A.~De Rujula, H.~Georgi, S.~L.~Glashow, {\em Phys.\ Rev.\ Lett.}\ {\bf 38}, 317 (1977).

\bibitem{Eichten80}
E.~Eichten, K.~Gottfried, T.~Kinoshita, K.~D.~Lane, Tung-Mow Yan,
{\em Phys.\ Rev.}\ {\bf D21}, 203 (1980).

\bibitem{Bradley}
A.~Bradley, D.~Robson, {\em Phys.\ Lett.}\ {\bf B93}, 69 (1980);
{\em Z.\ Phys.}\ {\bf C6}, 57 (1980).

\bibitem{lightMolecules}
R.~L.~Jaffe, {\em Phys.\ Rev.}\ {\bf D15}, 267 (1977);
J.~Weinstein, N.~Isgur, {\em Phys.\ Rev.\ Lett.}\ {\bf 48}, 659 (1982),
{\em Phys.\ Rev.}\ {\bf D27}, 588 (1982), {\em Phys.\ Rev.}\ {\bf D41}, 2236 (1990).

\bibitem{Tornqvist}
N.~A.~Tornqvist, {\em Phys.\ Rev.\ Lett.}\ {\bf 67}, 556 (1991), 
{\em Z.~Phys.}\ {\bf C61}, 525 (1994),
hep-ph/0308277.

\bibitem{otherPionEx}
A.~V.~Manohar, M.~B.~Wise, {\em Nucl.\ Phys.}\ {\bf B339}, 17 (1993);
T.~E.~O.~Ericson, G.~Karl, {\em Phys.\ Lett.}\ {\bf B309}, 426 (1993).

\bibitem{ClosePage}
F.~E.~Close, P.~R.~Page, hep-ex/0309253.

\bibitem{BraatenKusunoki}
E.~Braaten, M.~Kusunoki, hep-ph/0311147. 
 

\bibitem{my-mpp}
The data come from Fig.3a in BELLE's paper\rlap.\,\cite{BELLEX}
The sideband distribution was subtracted here.
The binning and normalization were also changed. 

\bibitem{BelleKpsipp}
K.~Abe \etal\ (BELLE), hep-ex/0307061. 

\bibitem{YuanQiaoChao97}
F.~Yuan, C.-F.~Qiao , K.-T.~Chao,  {\em Phys.\ Rev.}\ {\bf D56}, 329 (1997). 

\bibitem{EichtenLaneQuigg}
E.~J.~Eichten, K.~Lane, C.~Quigg, {\em Phys.\ Rev.\ Lett.}\ {\bf 89}, 162002 (2002).

\bibitem{EichtenQWG} 
E.~J.~Eichten,
at International Workshop on Heavy Quarkonium
at Fermilab, Batavia, IL, USA, September 20-22, 2003.


\bibitem{Fulcher91}
L.~P.~Fulcher, {\em Phys.\ Rev.}\ {\bf D44}, 2079 (1991).

\bibitem{p1d2pot}
The following sample of potential models
(ordered according to the increasing
value of $\sigma(\Delta M_{cog})$)
is displayed in Fig.~\protect\ref{fig:ccdmass}: %\hfill
H.~Ito, {\em Prog.\ of Theor.\ Phys.}\ {\bf 84}, 94  (1990);
D.~Ebert, R.~N.~Faustov, V.~O.~Galkin, {\em Phys.\ Rev.}\ {\bf D67}, 014027 (2003);
E.~Eichten, F.~Feinberg, {\em Phys.\ Rev.}\ {\bf D23}, 2724 (1981);
S.~Godfrey, N.~Isgur, {\em Phys.\ Rev.}\ {\bf D32}, 189 (1985);
T.~A.~Lahde, {\em Nucl.\ Phys.}\ {\bf A714}, 183 (2003);
N.~Barik, S.~N.~Jena, {\em Phys.\ Rev.}\ {\bf D24}, 680 (1981);
W.~Buchmuller, {\em Phys.\ Lett.}\ {\bf B112}, 479 (1982);
L.~P.~Fulcher, {\em Phys.\ Rev.}\ {\bf D44}, 2079 (1991);
J.~Baacke, Y.~Igarashi, G.Kasperidus, {\em Z.\ Phys.}\ {\bf C13}, 131 (1982);
J.~R.~Hiller, {\em Phys.\ Rev.}\ {\bf D30}, 1520 (1984);
M.~Hirano, {\em Prog. of Theor. Phys.}\ {\bf 83}, 575 (1990);
T.~A.~Lahde, C.~J.~Nyfalt, D.~O.~Riska, {\em Nucl.\ Phys.}\ {\bf A645}, 587 (1999).

\bibitem{BELLEPrivate}
BELLE collaboration, private communication.

\bibitem{PakvasaSuzuki}
S.~Pakvasa, M.~Suzuki, hep-ph/0309294.

\end{thebibliography}
\end{document}